\documentclass[12pt,a4paper]{article}
\usepackage[utf8]{inputenc}
\usepackage[longnamesfirst]{natbib}

\usepackage[top=36truemm,bottom=36truemm,left=27truemm,right=27truemm]{geometry}

\usepackage{amsthm, amssymb, amsmath, appendix}
\usepackage{hyperref}
\usepackage{mathrsfs}
\usepackage{physics}
\usepackage{mathtools, enumerate}

\usepackage{mathtools}
\mathtoolsset{showonlyrefs=true} 

\theoremstyle{definition}
\newtheorem{definition}{Definition}
\newtheorem{example}{Example}
\newtheorem{remark}{Remark}

\theoremstyle{theorem}
\newtheorem{theorem}{Theorem}

\newtheorem{proposition}{Proposition}
\newtheorem{corollary}{Corollary}
\newtheorem{claim}{Claim}

\newtheorem{axiom}{Axiom}
\newtheorem{axiomB}{Axiom$^\ast$}

\newcommand{\stsucc}{\succ\!\!\!\succ}

\usepackage{tikz}
\usetikzlibrary{intersections, calc, arrows.meta, positioning}
\usepackage{cancel}

\makeatletter 

\@addtoreset{figure}{section}
\@addtoreset{table}{section}
\makeatother 

\title{Preferences with Multiple Forecasts\thanks{The authors are grateful to Youichiro Higashi, Noriaki Kiguchi, Nobuo Koida, Hendrik Rommeswinkel, Taishi Sassono, Norio Takeoka, and Yusuke Yamaguchi for their helpful comments. The authors would like to thank participants of the Game Theory Workshop 2025 (Kanazawa, Japan). This research is supported by JSPS KAKENHI Grant Number 23KJ0979.}}
\author{Kensei Nakamura\thanks{Graduate School of Economics, Hitotsubashi University, Kunitachi, Tokyo 186-8601, Japan. E-mail: kensei.nakamura.econ@gmail.com}\hspace{1mm} and Shohei Yanagita\thanks{Graduate School of Economics, Hitotsubashi University, Kunitachi, Tokyo 186-8601, Japan. E-mail: shoheiyanagita@gmail.com}}
\date{\today}

\begin{document}

\maketitle

\vspace{7mm}

\begin{abstract}

When a collective decision maker presents a menu of uncertain prospects to her group members, each member’s choice depends on their predictions about payoff-relevant states.
In reality, however, these members hold different predictions; more precisely, they have different prior beliefs about states and predictions about the information they will receive.
In this paper, we develop an axiomatic framework to examine collective decision making under such disagreements.
First, we characterize two classes of representations: Bewley multiple learning (BML) representations, which are unanimity rules among predictions, and justifiable multiple learning (JML) representations, where a single prediction has veto power.
Furthermore, we characterize a general class of representations called hierarchical multiple learning representations, which includes BML and JML representations as special cases. Finally, motivated by the fact that these representations violate completeness or intransitivity due to multiple predictions, we propose a rationalization procedure for constructing complete and transitive preferences from them.

\vspace{3mm}
\end{abstract}

\newpage
\section{Introduction}

\sloppy

Collective decision makers (henceforth, DMs) often face decision problems where they must choose a set of options for their group members.
These DMs aim to present an appropriate menu while considering the members' choices in the future. 
In such cases, each member's choice from the menu depends on their own predictions about payoff-relevant states.
For instance, companies design savings plans or provide skill development opportunities for their employees, and the employees’ choices of a particular savings plan or skill development package would depend on their predictions about future economic conditions or technological trends.
Similarly, governments organize healthcare or employment support services for their citizen, and the citizen's choices from these services would be influenced by their predictions about future pandemic risks or labor market demands.

One of the challenges of these collective decisions is that predictions about states often vary among members.
While one member might be optimistic about future economic conditions, another might take a more pessimistic view.
As a result, optimistic and pessimistic members prefer different options.
Moreover, in many situations, including the example of predicting economic trends, each member updates their beliefs by observing signals and makes their choice based on the updated beliefs.
Therefore, not only differences in beliefs about the states but also differences in predictions about the information they will receive play a crucial role in their choices.
These two types of disagreement about predictions lead to differences in members’ choices even when they are offered the same menu.

Compared to disagreement in beliefs about states, disagreement in predictions about future information has received little attention in the literature.
To further illustrate the importance of this disagreement, consider the following example.
Two individuals, Alice and Bob, are looking for new jobs. The government provides them with employment support programs as menus of several job options. 
After a menu is presented, Alice and Bob choose one job from it. 
The benefits they obtain from each job (e.g., salary or employment duration) depend on the “demand for skills in the labor market” at the time of selection.
Let us suppose that Alice and Bob have an identical prior belief about labor market demand  but differing predictions about future information. 
Alice expects to gain detailed information about the labor demand in the future and prefers a menu that allows her to select the most suitable job based on the information that she will obtain. 
Bob, on the other hand, does not expect to receive enough information in the future and prefers a menu that includes stable jobs unaffected by potential changes in the labor market demand. That is, Bob is less likely to prefer a menu that consists of jobs that offer high benefits only if a specific demand is realized.
As a result, although they share a common prior belief, they make different choices due to disagreement in predictions about future information.

Under such disagreement, how should the government evaluate menus? In this paper, we examine these collective decision problems from the axiomatic perspective. More precisely, this paper models several decision criteria for menus and establishes their axiomatic foundations. By axiomatizing these criteria, we clarify the behavioral foundations underlying each criterion.  The next subsection provides an overview of our framework and results.

\subsection{Results overview}

In this paper, we study a DM's preference relation over menus.
A menu is formalized as a finite set of acts, where an act $f$ is a function from $\Omega$ to $X$.
Here, $\Omega$ is a state space, and $X$ is the set of lotteries over deterministic prizes.
We consider the following decision-making flow:

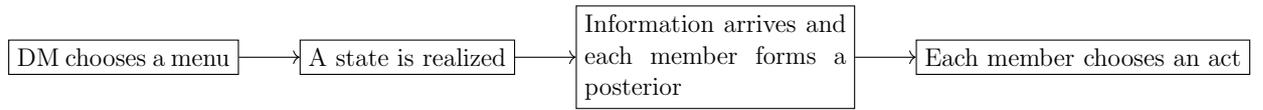
\begin{center}
\begin{tikzpicture}[node distance=1 cm, every node/.style={font=\normalsize, draw, rectangle}, scale=0.8, transform shape]

    \node (step1) {\parbox{3.5 cm}{DM chooses a menu}};
    \node (step2) [right=of step1] {A state is realized};
    \node (step3) [right=of step2] {\parbox{4.3 cm}{Information arrives and each member forms a posterior}};
    \node (step4) [right=of step3] {Each member chooses an act};

    \draw[->] (step1) -- (step2);
    \draw[->] (step2) -- (step3);
    \draw[->] (step3) -- (step4);

\end{tikzpicture}
\end{center}

\noindent
First, the DM chooses a menu $F$.
Then, each member observes information,  which arises probabilistically depending on the realized state, and forms a posterior belief by applying Bayes' rule to her prior belief.
Finally, based on that posterior belief, each member chooses an act from the menu $F$ and then recognizes the realized state and the outcome she obtains.
Since members have prior beliefs and predict information that they will obtain, each member's prediction can be represented as an element of $\Delta(\Delta(\Omega))$, a probability distribution over posterior beliefs.
In the case where the DM consists of multiple members, the aforementioned “disagreement over predictions” can be represented as  different  $\pi, \pi'\in\Delta(\Delta(\Omega))$.
We refer to each element in $\Delta(\Delta (\Omega))$ as an information structure.

To introduce the definition of decision criteria studied in this paper, we define the value of a menu with respect to an information structure.
For a utility function $u$ and a menu $F$, the benefit of information of $\pi$ is defined as follows: 
\begin{equation*}
    b^u_F (\pi) = \int_{\Delta (\Omega)} \qty[ \max_{f\in F} \qty( \int_\Omega u(f(\omega)) p (d\omega) ) ] \pi (dp). 
\end{equation*}
This represents the expected utility value of a member who anticipates having each posterior belief $p$ with probability $\pi(p)$ and chooses the most preferred act from $F$ after receiving information. 
The max part corresponds to the benefit of $F$ under a posterior belief $p$. 
By taking its expectation according to her information structure $\pi$, $b^u_F (\pi)$ represents the expected value of $F$ before obtaining information from $\pi$. 

Given this definition, we first propose a unanimity rule and provide its axiomatic characterization.
We say that a DM with a preference $\succsim$ admits a \textit{Bewley multiple learning (BML) representation} if for some $u:X\rightarrow\mathbb{R}$ and $\Pi\subset\Delta(\Delta(\Omega))$, 
\[
F \succsim G \iff \min_{\pi\in\Pi}(b^u_F (\pi)-b^u_G (\pi))\geq 0
\]
for all menus $F$ and $G$. 
According to this criterion, a menu $F$ is preferred to another menu $G$ if and only if, for any $\pi\in\Pi$, the expected value of $F$ is higher than that of $G$.
However, since this preference requires a unanimous agreement among all predictions in $\Pi$, it does not satisfy completeness.
On the other hand, the second class of criteria is a natural opposite counterpart to the BML preferences. 
We say that a DM with a preference $\succsim$ admits a \textit{justifiable multiple learning (JML) representation} if for some $u:X\rightarrow\mathbb{R}$ and $\Pi\subset\Delta(\Delta(\Omega))$, 
\[
F \succsim G \iff \max_{\pi\in\Pi}(b^u_F (\pi)-b^u_G (\pi))\geq 0
\]
for all  menus $F$ and $G$.
That is, $F$ is preferred to $G$ under a JML preference if and only if there exists at least one $\pi\in\Pi$ where the benefit of $F$ is higher than that of $G$.
According to this preference, each member can be interpreted as having  veto power.
If one member’s prediction suggests that $F$ is better than $G$, even if all of the other members evaluate them in the opposite direction, $F$ is considered as good as $G$.
As a result, JML preferences satisfy completeness but not transitivity.

We characterize each class of decision criteria by imposing relaxed rationality axioms together with several basic axioms. The limitation of BML preferences is the lack of decisiveness under uncertainty, and this violation of rationality can capture the key behavioral implication of these preferences. 
JML preferences satisfy only a weak version of transitivity in general, and an axiom motivated by this observation can characterize these preferences. 
Notice that these irrationalities stem from the presence of distinct predictions. We show that the comparative statics of sets of information structures in both classes can be characterized by comparing the degree to which rationality is violated.

In the latter part of this paper, we introduce a general class of criteria that includes both BML and JML preferences as special cases and provide its axiomatic characterization.
Such general preferences are represented using a collection of sets of information structures to capture other ways of   aggregating the members' opinions.
We call such a criterion a \textit{hierarchical multiple learning (HML) preference} and define it as follows: For all  menus $F$ and $G$, 
\[
F \succsim G \iff \max_{\Pi \in \mathbf{\Pi}} \min_{\pi \in \Pi} (b_F^u (\pi) -  b_G^u (\pi)) \geq 0.
\]
Here, $\boldsymbol{\Pi}$ is a collection of subsets of $\Delta(\Delta(\Omega))$, and each $\Pi\in\boldsymbol{\Pi}$ represents predictions of a group of members.
According to this criterion, if there exists a group of members that ranks $F$ higher than $G$ unanimously, then $F$ is preferred to $G$.
If there is only one group, it corresponds to a BML preference, and if each group consists of a single individual, it corresponds to a JML preference.
We show that these preferences with hierarchical structures can be characterized using axioms shared by the BML and JML preferences.

BML, JML, and HML preferences are irrational decision criteria because they fail to satisfy completeness or transitivity.
In collective decision makings, however, it would be undesirable that decision criteria do not satisfy completeness or transitivity.
To bridge this gap, we propose a procedure that constructs complete and transitive preferences from  the preferences mentioned above.
These procedures offer a theoretically founded guideline on how to “practically” refer to the different predictions of members.

\subsection{Related literature}

All of the decision-making criteria studied in this paper can be considered generalizations of  subjective learning (SL) preferences axiomatized by \citet{dillenberger2014theory}.
A DM with an SL preference has one information structure $\pi$ in her mind and evaluates each menu $F$ according to its benefit of information $b^{u}_{F}(\pi)$ of $\pi$. 
Thus, any BML or JML preference is reduced to an SL preference when the set $\Pi$ of information structures is a singleton.
Similarly, an HML preference  with $\mathbf{\Pi}$ coincides with an SL preference when $\mathbf{\Pi} = \{\{\pi\}\}$ for some information structure $\pi$.
While SL preferences focus on individual learning behavior, our motivation lies in studying collective DM who faces disagreements about future predictions.

Several extensions of SL preferences have been considered. 
\citet{de2017rationally} developed a model of costly information acquisition.
In their model, the DM chooses an optimal information structure compatible with her prior while taking information costs into account.
\citet{pennesi2015costly} also developed a similar model in another domain. 
\citet{higashi2020wpCost} extended  \citeauthor{de2017rationally}'s (\citeyear{de2017rationally}) model to accommodate more general cost structure. 
In addition, \citet{higashi2023wpHomo} studied special cases of \citet{higashi2020wpCost}. 
\citet{higashi2023wpCorse} examined a generalization of the SL model, where the DM does not necessarily choose information in an optimistic way, but rather may evaluate them cautiously.
As another direction, \citet{pennesi2024subjective} developed a model in which a DM subjectively predicts the timing of information arrivals.
All of these models were characterized by weakening the axioms of independence or preference for flexibility in the characterization of  SL preferences.
In contrast, we focus on the violation of completeness and transitivity and weaken these axioms. This approach characterizes BML, JML, and HML preferences, which accommodate collective decision-making rules rather than an individual DM, as generalizations of SL preferences.

 In addition, our research is closely related to studies on incomplete or intransitive preferences in \citeauthor{anscombe1963definition}'s (\citeyear{anscombe1963definition}) framework.\footnote{More precisely, this framework was first introduced by \citet{anscombe1963definition} and elaborated by \citet{fishburn1970book}. }
The most related models are the ones proposed by \citet{bewley2002knightian} and \citet{lehrer2011justifiable}. 
\citet{bewley2002knightian} introduced an incomplete decision-making model with multiple priors. Precisely, a DM with some Bewley preference thinks one act $f$ to be better than another $g$ if and only if for all priors in her mind, expected utility of $f$ is larger than that of $g$. 
On the other hand, \citet{lehrer2011justifiable} examined preferences with multiple priors but used them in the opposite way.
These preferences are called the justifiable preferences. A DM with a justifiable preference thinks  $f$ to be better than $g$ if and only if for at least one prior in her mind, expected utility of $f$ is larger than than that of $g$. 
These models can capture disagreements of priors but not disagreements of predictions of future information. 
Hence, this paper can be considered as extending these findings to the context of information acquisition and proposing preferences that accommodate these two disagreements.
Furthermore, \citet{lehrer2011justifiable} considered the counterpart of HML preferences as well.\footnote{
\citet{chandrasekher2022dual} and \citet{Xia2020wp} also considered complete and transitive preferences admitting representations with the max-of-min operators. } 

Other than these two models, incomplete and/or intransitive preferences have been considered under risk (e.g., \citet{baucells2008multiperson,dubra2004expected,hara2019coalitional, kochov2020subjective}) and uncertainty (e.g.,  \citet{nau1992indeterminate,galaabaatar2013subjective,ok2012incomplete,nascimento2011class,faro2015variational,echenique2022twofold}). 
Particularly, \citet{kochov2020subjective} characterized incomplete preferences that evaluate menus of lotteries based on the unanimity rule with respect to multiple probability distributions over a subjective state space.
While Kochov's model represents disagreements about beliefs over subjective states, our model deals with disagreements about beliefs and predictions of future information given an objective state space. 

In a very general setting, \citet{evren2011multi} showed that incomplete but transitive preference can be represented as the unanimous rule among multiple utility functions. \citet{nishimura2016utility} examined various classes and found that (i) any complete but intransitive preference has a structure similar to the JML preferences and (ii)  any reflexive preference has a structure similar to the HML preferences. 
Our characterization results identifies the condition that disagreements stem from only predictions, not tastes or correlations between them, at the axiomatic level.

Finally, we briefly discuss the literature related to procedures of constructing rational preferences from irrational ones. After \citet{gilboa2010objective} introduced this problem, many papers considered their generalizations (e.g., \citet{cerreia2016objective,cerreia2020rational,danan2016robust,echenique2022twofold,faro2015variational,frick2022objective,grant2020objective,kopylov2009choice}) or extensions to dynamic setup (e.g., \citet{faro2019dynamic,bastianello2022dynamically}) and risky situations (e.g., \citet{cerreia2015cautious}).
Most papers have focused on the rationalization of incomplete but transitive preferences such as the Bewley preferences. 
On the other hand, we consider the rationalization procedure of the classes that we characterize, and hence deal with not only incomplete but also intransitive preferences. 
Note that \citet{nakamura2025cautious} considered the rationalization procedure of incomplete and intransitive preferences with hierarchical structures in the \citeauthor{anscombe1963definition}'s framework.
The paper proposed a cautious way to construct rational preferences to provide a normative justification for the proposed decision-making model. 
On the other hand, this paper characterizes a more general class of procedures, including cautious ways and optimistic ones.

\subsection{Outline of the paper}

The rest of this paper is organized as follows: 
Section 2 provides our framework and the formal definitions of BML and JML preferences. 
Section 3 establishes axiomatic characterizations of these decision criteria. 
Section 4 examines the comparative statics of the parameters in BML and JML preferences. 
Section 5 introduces general decision criteria, or HML preferences, and provides its axiomatic characterization.
Section 6 proposes a procedure for constructing complete and transitive preferences based on the above decision criteria.

\section{Multi-Learning Preferences}

In this section, we introduce the preliminary setup and provide definitions of BML and JML preferences.

\subsection{Framework}

Our framework is based on \citeauthor{de2017rationally}'s (\citeyear{de2017rationally}) model. Let $\Omega$ be a finite set of states and $X$ be a set of outcomes, consisting of lotteries over deterministic prizes such as monetary payoff. 
An \textit{act} is a function $f: \Omega \rightarrow X$, and the set of acts is denoted by $\mathcal{F}$.  
A \textit{menu} is a finite set of acts and the set of menus is denoted by $\mathbb{F}$. Typical elements of $\mathbb{F}$ are denoted by $F$,  $G$,  or $H$. 
With some abuse of notation, we identify an outcome $x\in X$ with the constant act $f$ such that $f(\omega) = x$ for all $\omega \in \Omega$. 
Similarly, we identify an act $f \in \mathcal{F}$ with the singleton menu $\{ f \} \in \mathbb{F}$. 

A DM has a binary relation $\succsim$ over $\mathbb{F}$.
Since we consider a collective DM whose members choose an act from a given menu after they receive information, $F\succsim G$ should be interpreted that the DM weakly prefers $F$ to $G$ as her members' opportunity set at the time before they receive information. 
The asymmetric and  symmetric parts of $\succsim$ are denoted by $\succ$ and $\sim$, respectively. 

For $f,g \in \mathcal{F}$ and $\alpha \in [0,1]$, let $\alpha f + (1 - \alpha) g$ be the act $h$ such that for all $\omega \in \Omega$, $h(\omega) =  \alpha f(\omega) +(1- \alpha) g(\omega)$. 
Similarly, for $F,G \in \mathbb{F}$ and $\alpha \in [0,1]$, let $\alpha F + (1 - \alpha) G$ be the menu $H$ such that
\begin{equation*}
    H = \{ \alpha f + (1 - \alpha) g : f\in F ~ \text{and} ~ g\in G \}. 
\end{equation*}
The assumption that $X$ consists of lotteries is needed to define these mixture operations.

Let $\Delta (\Omega)$ be the set of probability distributions over $\Omega$.
Since $\Omega$ is finite, we endow $\Delta(\Omega)$ with the topology induced by the Euclidean distance.
Its typical elements are denoted by $p$ or $p'$. 
Let $\Delta (\Delta (\Omega))$ be the set of Borel measurable probability distributions over $\Delta (\Omega)$. 
We endow $\Delta(\Delta(\Omega))$ with the weak* topology.\footnote{For a more precise definition, see Appendix A.1.}
Its typical elements are denoted by $\pi$ or $\pi'$. 
Each element $\pi$ can be interpreted as a prediction about posterior distributions over states.
If a member holds the prediction  $\pi$, it can be interpreted that she believes her posterior belief will be  $p$  with probability $ \pi(p)$.
Note that her prior belief is obtained from $\pi$ by calculating the expectation $\mathbb{E}_{\pi}[p]$.

\subsection{Preferences with multiple information structures}

\citet{dillenberger2014theory} considered DMs who expect to update a prior belief over states according to an information structure $\pi \in \Delta (\Delta (\Omega))$.
Given $F\in \mathbb{F}$ and a utility function $u: X\rightarrow \mathbb{R}$, the \textit{benefit of information} of $\pi \in \Delta (\Delta (\Omega))$ is 
\begin{equation*}
    b^u_F (\pi) = \int_{\Delta (\Omega)} \qty[ \max_{f\in F} \qty( \int_\Omega u(f(\omega)) p (d\omega) ) ] \pi (dp).  
\end{equation*}
This represents the expected utility value of an agent who anticipates having a posterior belief $p$ with probability $\pi(p)$ and chooses the most preferred act from $F$ after receiving information.

\citet{dillenberger2014theory} introduced an SL representation, which is formally defined as follows: A preference relation $\succsim$ on $\mathbb{F}$ admits a \textit{subjective learning (SL) representation} if there exist a nonconstant affine function $u:X\rightarrow\mathbb{R}$ and $\pi\in\Delta(\Delta(\Omega))$ such that for all $F, G\in \mathbb{F}$, 
\begin{equation*}
    F\succsim G \iff b^{u}_{F}(\pi)\geq b^{u}_{G}(\pi).\footnote{
    More precisely, \citet{dillenberger2014theory} considered an act as a mapping $f:\Omega\rightarrow[0,1]$. In Appendix A of \citet{dillenberger2014theory}, they introduced the Anscombe-Aumann setting and derived the same result as that in the main text.}
\end{equation*}
In this representation, the DM utilize only one information structure $\pi$ when evaluating menus. 

While keeping the feature that the DM evaluates menus based on the benefit of information, we extend  this representation to consider cases where members in a group have different predictions about posterior distributions. 

The first representation we study is as follows.

\begin{definition}
    For a nonconstant affine function $u: X \rightarrow \mathbb{R}$ and a nonempty closed convex set $\Pi \subset \Delta (\Delta (\Omega ))$, 
    a binary relation $\succsim$ over menus admits a \textbf{\textit{Bewley multiple learning (BML) representation}} $(u, \Pi)$
    if   for all $F, G \in \mathbb{F}$, 
    \begin{equation}
    \label{eq_defBML}
        F \succsim G \iff \qty[ ~ b^u_F (\pi) \geq b^u_G (\pi) ~~ \text{for all $\pi \in \Pi$} ~ ].
    \end{equation}
\end{definition}

\noindent
When we do not need to mention the parameters $u$ and $\Pi$, we simply say that $\succsim$ admits a BML representation. 
Moreover, we sometimes refer to a preference that admits a BML representation as a BML preference. 
We use this terminology for the representations that we define later (Definitions \ref{def:jml} and \ref{def:hml}) without making specific remarks. 

The BML preferences are named after Bewley preferences (\citet{bewley2002knightian}).
A DM with a BML preference (\ref{eq_defBML}) ranks $F$ to be better than $G$  if $F$ attains a higher expected benefit than $G$ for every prediction $\pi\in \Pi$. 
If any disagreement happens among $\Pi$ when evaluating two menus, the DM does not make a decision.
In this sense, BML preferences behave according to the unanimity rule within the associated information structures.
Notice that the evaluation of lotteries is calculated using the same vNM function $u$.

\begin{example}[Incompleteness of BML preferences]
    Consider a binary state space $\Omega=\{\omega_{1},\omega_{2}\}$.
    Let $p\in\Delta(\Omega)$ be a probability distribution such that $p(\omega_{1})=p(\omega_{2})=\frac{1}{2}$ and $\pi\in\Delta(\Delta(\Omega))$ be an information structure such that $\pi(\delta_{\omega_{1}})=\frac{1}{2}$ and $\pi(\delta_{\omega_{2}})=\frac{1}{2}$, where $\delta_{\omega}$ is the probability measure $q$ such that $q(\omega)=1$.
    Suppose that a DM has a BML preference $(u,\Pi)$ such that $\Pi=\mathrm{conv}\{\delta_{p},\pi\}=\{\alpha\delta_{p}+(1-\alpha)\pi\in\Delta(\Delta(\Omega))\mid \alpha\in[0,1]\}$.\footnote{For a set $A$, let $\mathrm{conv}\, A$ denote the convex hull of $A$.}
    Consider the following two menus: 
    One is a singleton $\{f\}$ such that $u(f(\omega_{1}))=u(f(\omega_{2}))=2$, and
    the other consists of two acts $g$ and $h$ such that $u(g(\omega_{1}))=u(h(\omega_{2}))=3$ and $u(g(\omega_{2}))=u(h(\omega_{1}))=0$.
    Then, for $\delta_{p}$, $b^{u}_{\{f\}}(\delta_{p})=\frac{1}{2}\cdot2+\frac{1}{2}\cdot2=2>b^{u}_{\{g,h\}}(\delta_{p})=\frac{1}{2}\cdot3+\frac{1}{2}\cdot0=\frac{3}{2}$.
    On the other hand, for $\pi$, $b^{u}_{\{f\}}(\pi)=\frac{1}{2}\cdot2+\frac{1}{2}\cdot2=2<b^{u}_{\{g,h\}}(\pi)=\frac{1}{2}\cdot3+\frac{1}{2}\cdot3=3$.
    Thus, under this BML preference, $\{f\}$ and $\{g,h\}$ are incomparable. 
\end{example}

Note that the right hand side of 
\eqref{eq_defBML} can be rewritten as 
\begin{equation*}
    \min_{\pi \in \Pi} ( b^u_F (\pi) - b^u_G (\pi)) \geq 0. 
\end{equation*}
We sometimes use this inequality as the definition of the BML representations.

Next, we model DMs with multiple information structures but utilize them in an \textit{optimistic} way.
More precisely, we consider the DM who prefers a menu $F$ to a menu $G$ if and only if at least one information structure justifies such an evaluation.
The following is the formal definition of this  decision criterion:
\begin{definition}
\label{def:jml}
    For a nonconstant affine function $u: X \rightarrow \mathbb{R}$ and a nonempty closed convex set $\Pi \subset \Delta (\Delta (\Omega ))$, 
    a binary relation $\succsim$ over menus admits a \textbf{\textit{justifiable multiple learning (JML) representation}} $(u,\Pi)$
    if   for all $F, G \in \mathbb{F}$, 
    \begin{equation}\label{eq_JMLP}
        F\succsim G \iff \qty[~ \exists\pi\in\Pi\:\mathrm{s.t.}\:b^{u}_{F}(\pi)-b^{u}_{G}(\pi)\geq 0 ~ ].
    \end{equation}
\end{definition}

\noindent
The JML preferences are named after justifiable preferences characterized by \citet{lehrer2011justifiable} in the Anscombe-Aumann framework.
According to (\ref{eq_JMLP}), $F$ is weakly preferred to $G$ if and only if $F$ attains a higher expected benefit than $G$ for some information structure $\pi\in\Pi$.
Note that under this decision criterion, any disagreement within $\Pi$ is treated as an indifference relation.
Therefore, the JML preferences can be viewed as an aggregation process among members with veto power. Since any disagreement results in indifference relations, the JML preferences form complete preference relations.
However, they do not necessarily satisfy transitivity.
Below is a simple numerical example that demonstrates the violation of transitivity.

\begin{example}[Intransitivity of JML preferences]
    Consider a binary state space and a DM with a JML preference $\Pi$, which is the same as that of Example 1.
    In addition to $\{f\}$ and $\{g,h\}$ introduced in Example 1, consider another singleton $\{f^{*}\}$ such that $u(f^{*}(\omega_{1}))=u(f^{*}(\omega_{2}))=\frac{5}{2}$.
    Then, for both $\delta_{p}$ and $\pi$, $b^{u}_{\{f^{*}\}}(\delta_{p})=b^{u}_{\{f^{*}\}}(\pi)=\frac{5}{2}$ holds.
    Since $b^{u}_{\{f^{*}\}}(\delta_{p})=2.5>b^{u}_{\{g,h\}}(\delta_{p})=\frac{3}{2}$ and $b^{u}_{\{f^{*}\}}(\pi)=2.5<b^{u}_{\{g,h\}}(\pi)=3$, $\{f^{*}\}\sim \{g,h\}$ holds.
    Moreover, by the calculation in Example 1, $\{f\}\sim\{g,h\}$ holds.
    However, since $b^{u}_{\{f^{*}\}}(\delta_{p})=b^{u}_{\{f^{*}\}}(\pi)=\frac{5}{2}>b^{u}_{\{f\}}(\delta_{p})=b^{u}_{\{f\}}(\pi)=2$, $\{f^{*}\}\succ\{f\}$ holds, which violates transitivity.
\end{example}

As the BML preferences, the right hand side of \eqref{eq_JMLP} can be rewritten as
\begin{equation*}
     \max_{\pi \in \Pi} ( b^u_F (\pi) - b^u_G (\pi)) \geq 0.
\end{equation*}
Again, we sometimes use this inequality as the definition of the JML representations. 

\subsubsection{Relation to the Bewley and justifiable preferences}

In the Anscombe-Aumann framework, \citet{bewley2002knightian} characterized the following representations: For all $f,g\in \mathcal{F}$, 
\begin{equation*}
    f\succsim g\iff \qty[ ~ \int u(f(\omega))p(d\omega) \geq  \int u(g(\omega))p(d\omega) ~~ \text{for all $p \in P$} ~ ],
\end{equation*}
where $P\subset\Delta(\Omega)$ is a nonempty closed convex set.
These preferences are called Bewley preferences.
Similarly, \citet{lehrer2011justifiable} characterized an optimistic counterpart of Bewley preferences: For all $f,g\in \mathcal{F}$
\begin{equation*}
    f \succsim g \iff \qty[ ~  \exists p\in P \:\mathrm{s.t.} \int u(f(\omega))p(d\omega) \geq  \int u(g(\omega))p(d\omega) ~ ].
\end{equation*}
These preferences are called justifiable preferences.
Since these preferences evaluate acts unanimously or optimistically based on multiple beliefs, they are similar to BML and JML preferences, respectively.
The difference between these preferences and ours is that while Bewley and justifiable preferences only capture disagreement in beliefs about states, our models also capture disagreement in predictions about how those beliefs are updated---that is, what information can be obtained in the future.
This additional structure is possible because we introduce a dynamic decision timeline. 
To capture disagreements in predictions about future information, it is necessary to consider the timeline over which the members can utilize these information.

Notice that the BML and JML preferences do not exclude the possibility of disagreement in beliefs about states.
In fact, on singletons, the BML and JML preferences behave in the same way as Bewley and justifiable preferences, respectively.
For a BML preference $(u,\Pi)$,
\begin{equation*}
    \{f\}\succsim \{g\}\iff \qty[~  \int u(f(\omega))\mathbb{E}_{\pi}[p](d\omega) \geq  \int u(g(\omega))\mathbb{E}_{\pi}[p](d\omega) ~~ \text{for all $\pi\in\Pi$} ~ ]
\end{equation*}
holds.
Similarly, for a JML preference $(u,\Pi)$,
\begin{equation*}
    \{f\}\succsim \{g\}\iff \qty[~ \exists\pi\in\Pi\:\mathrm{s.t.}\:\int u(f(\omega))\mathbb{E}_{\pi}[p](d\omega) \geq  \int u(g(\omega))\mathbb{E}_{\pi}[p](d\omega) ~ ]
\end{equation*}
holds.
In this sense, we can interpret the BML and JML preferences as generalizations of Bewley preferences and justifiable preferences, respectively.

\section{Representation Theorems}

This section provides axiomatic characterizations of the BML and JML preferences, clarifying the differences between them at the level of observable behaviors. 

\subsection{BML preferences}

First, we provide the characterization of the BML preferences. 
The first axiom postulates that the DM's preference $\succsim$ is not trivial. 

\begin{axiom}[Nontriviality]
    There exist $F, G \in \mathbb{F}$ with $F \succ G$.
\end{axiom}

Next, we introduce axioms of rationality. 
The following axiom states that when comparing two lotteries, the DM can determine which is the preferred one, including a tie. That is, indecisiveness does not stem from tastes for options with no uncertainty. 

\begin{axiom}[Completeness for lotteries]
    For all $x, y \in X$, $x\succsim y$ or $y\succsim x$.\footnote{Note that our characterization result can be obtained if we strengthen \textit{completeness for lotteries} as follows: For all $F,G\in\mathbb{F}$ such that $F,G \subset X$, $F\succsim G$ or $G\succsim F$.} 
\end{axiom}

Since the DM with a BML preference cannot always determine which of two menus is better due to disagreements among members, we consider the above weak completeness.

The following axiom is the standard axiom of transitivity.

\begin{axiom}[Transitivity]
\label{axiom:trans}
     For all $F, G, H \in \mathbb{F}$, if  $F\succsim G$ and $G\succsim H$, then $F\succsim H$. 
\end{axiom}

While \textit{transitivity} is a standard requirement in economics, it should be noted that it is not entirely innocent.
For instance, when considering a collective DM, this axiom excludes decision criteria such as the plurality rule.
It appears reasonable for a group with disagreements about predictions to evaluate menus through the plurality rule among its members.
However, in this subsection, we exclude these possibilities by focusing on preferences that satisfy \textit{transitivity}.
We will explore a weaker form of \textit{transitivity} and related decision criteria in the next subsection.

Then, we introduce an axiom of continuity. While our continuity axiom is stronger than those introduced in \citet{dillenberger2014theory} and \citet{de2017rationally}, the underlying intuition remains unchanged.\footnote{
For the definition of their continuity axiom, see \textit{mixture continuity} (Axiom \ref{axiom:mixcon}). 
} 

\begin{axiom}[Continuity]
\label{axiom:cont}
    The set $\{ \alpha \in [0,1] : \alpha F + (1- \alpha) F' \succsim \alpha G + (1- \alpha) G' \}$ is  closed for all $F, F', G, G' \in \mathbb{F}$.\footnote{A similar axiom is discussed in \citet{dubra2004expected} under risk. They characterized a class of incomplete preferences called expected multi-utility preferences using this continuity axiom.}
\end{axiom}

According to the decision-making flow  we consider, each member chooses the best act from a given menu after obtaining information. Therefore, the DM should prefer larger menus since these menus allow the members to make choices more flexibly.
This property was first formalized by \citet{kreps1979representation} and is defined as follows: 

\begin{axiom}[Preference for flexibility]
\label{axiom:p4flex}
    For all $F, G\in \mathbb{F}$, if $G\subset F$, then $F\succsim G$. 
\end{axiom}

Furthermore, since unambiguously worse acts are not chosen by the members regardless of the information they obtain, it is not worth adding such acts to menus. 
The next axiom states that if we add an act $g$ that is state-wise dominated by some act in the original menu $F$, then adding $g$ to $F$ does not alter the evaluation of $F$---that is, $F\cup \{ g \}$ is indifferent to $F$.

Before stating the axiom, we introduce a notation for the state-wise dominance between menus. 
For $F,G\in \mathbb{F}$, we state that $F$ \textit{state-wise dominates} $G$, denoted by $F\trianglerighteq_{D}G$, if for any $g\in G$, there exists $f\in F$ such that $f(\omega)\succsim g(\omega)$ for all $\omega\in\Omega$.

\begin{axiom}[Dominance]
\label{axiom:dominance}
    For all $F\in \mathbb{F}$ and $g\in \mathcal{F}$, if $F\trianglerighteq_{D} \{g\}$, then $F\sim F\cup \{ g \}$.
\end{axiom}

The following axiom is the standard independence axiom. For a detailed discussion on this axiom, see \citet{dillenberger2014theory}.

\begin{axiom}[Independence]
\label{axiom:ind}
    For all $F, G, H\in \mathbb{F}$ and $\alpha \in (0,1)$, 
    \begin{equation*}
        F\succsim G\iff \alpha F + (1 - \alpha) H \succsim \alpha G + (1 - \alpha) H. 
    \end{equation*}
\end{axiom}

The final axiom states that the DM is indifferent between a menu $F$ and its ex-post randomization.

\begin{axiom}[Indifference to ex-post randomization]
    \label{axm_rand}
    For all $F\in\mathbb{F}$, $n\in\mathbb{N}$, and $\beta_1,...,\beta_n\in[0,1]$ such that $\sum_{i=1}^n \beta_i=1$, 
    \begin{equation*}
        F\sim \sum_{i=1}^n \beta_i F.
    \end{equation*}
\end{axiom}

\textit{Indifference to ex-post randomization} was introduced in \citet{higashi2023wpCorse} in this framework, and a similar axiom was examined in \citet{dekel2001representing}, which studied preferences over menus of lotteries.  
If the members choose acts to maximize an expected utility, it can be justified.
In such cases, the members always make ex-post choices from the extreme points of $F$, and thus, the additional flexibility provided by ex-post randomization does not increase the menu’s value.
Therefore, $\sum_{i=1}^n \beta_i F$ becomes indifferent to $F$.
Notice that \textit{independence} does not imply this axiom.
This is because by the definition of mixture between menus, for any non-singleton menu $F$ and $\alpha\in(0,1)$, $F$ is not equal to $\alpha F+(1-\alpha)F$.


Henceforth, we refer to the set of axioms---\textit{nontriviality}, \textit{continuity}, \textit{independence}, and \textit{indifference to ex-post randomization}---as \textbf{basic axioms}. 
Since we use these axioms in all the characterization results, this labeling helps clarify the differences among the preferences studied in this paper.

Then, we provide a characterization result for the BML preferences. 
The following theorem shows that the axioms introduced in this subsection characterize the BML preferences. 
Furthermore, the uniqueness of parameters can be obtained. 

\begin{theorem}
\label{thm:BML}
    A binary relation $\succsim$ over menus  admits a BML representation $(u, \Pi)$ if and only if it satisfies\textit{ basic axioms}, \textit{completeness for lotteries}, \textit{transitivity}, \textit{preference for flexibility}, and \textit{dominance}.  

    Furthermore, if $\succsim$ admits another BML representation $(u', \Pi')$, then $u' = \alpha u + \beta$ for some $(\alpha, \beta ) \in \mathbb{R}_{++} \times \mathbb{R}$ and $\Pi = \Pi'$. 
\end{theorem}

\subsection{JML preferences}

In this subsection, we provide an axiomatic characterization of the JML preferences.
The first axiom is the standard completeness axiom.

\begin{axiom}[Completeness]
\label{axm_comp}
     For any $F,G\in\mathbb{F}$, $F\succsim G$ or $G\succsim F$.
\end{axiom}

\noindent
A DM with a JML preference considers a menu $F$ to be weakly better than a menu $G$ if at least one member believes so.
Therefore, as long as she believes that all members have complete preferences, this axiom would naturally be satisfied.

While we introduce a stronger axiom of completeness than the one imposed in Theorem \ref{thm:BML}, we consider a weaker axiom of transitivity. 
For menus $F$, $G$, and $H$, if $F$ is preferred to $G$ and $G$ state-wise dominates $H$ (or if $F$ state-wise dominates $G$ and $G$ is preferred to $H$), then $F$ is preferred to $H$. In other words, it requires the transitive property only when one menu is better than another under any prediction.

\begin{axiom}[Unambiguous transitivity]
\label{axm_Unambig_trans}
    For any $F,G,H\in\mathbb{F}$,
    \begin{enumerate}
        \item if $F\trianglerighteq_{D}G$ and $G\succsim H$, then $F\succsim H$; and
        \item if $F\succsim G$ and $G\trianglerighteq_{D}H$, then $F\succsim H$. 
    \end{enumerate}
\end{axiom}

While \textit{unambiguous transitivity} is novel in menu-preference literature, similar axioms have been presented in several papers.
For instance, \cite{lehrer2011justifiable} introduced a counterpart axiom in the Anscombe-Aumann framework, which states that for all acts $f,g,h\in\mathcal{F}$, (i) $f\trianglerighteq_{D}g$ and $g\succsim h$ imply $f\succsim h$, and (ii) $f\succsim g$ and $g\trianglerighteq_{D} h$ imply $f\succsim h$.
Since the set $\mathcal{F}$ of acts is a subdomain of the set $\mathbb{F}$ of menus, it is straightforward to see that \textit{unambiguous transitivity} implies this axiom.
Additionally, \citet{nau1992indeterminate} introduced an axiom similar to the first part of \textit{unambiguous transitivity}.


The final axiom requires that if a menu $F$ is strictly better than a menu $G$ and a menu $H$ is strictly better than a menu $H'$, then any mixture of better menus $F$ and $H$ is always preferred to that of worse menus $G$ and $H'$ with the same proportion.

\begin{axiom}[Favorable mixing monotonicity]
    \label{axm_mon}
    For any $F,G,H,H'\in\mathbb{F}$ and $\alpha\in[0,1]$, if $F\succ G$ and $H\succ H'$, then $\alpha F+(1-\alpha)H\succ\alpha G+(1-\alpha)H'$.
\end{axiom}

Note that if $\succsim$ satisfies \textit{transitivity}, then \textit{independence} implies \textit{favorable mixing monotonicity}.\footnote{To see this, let $\alpha\in[0,1]$ and $F,G,H,H'\in\mathbb{F}$ with $F\succ G$ and $H\succ H'$.
By \textit{independence}, $\alpha F+(1-\alpha)H\succ\alpha G+(1-\alpha)H$ and $\alpha G+(1-\alpha)H\succ\alpha G+(1-\alpha)H'$. 
By \textit{transitivity}, $\alpha F+(1-\alpha)H\succ\alpha G+(1-\alpha)H'$.} 
This is why we do not need to impose this axiom when characterizing BML preferences.
Since we do not impose \textit{transitivity} in this subsection, it is necessary to incorporate this axiom when characterizing JML preferences. 

The following result shows that under the above axioms combined with \textit{basic axioms}, we can provide a characterization of JML preferences.
Furthermore, as Theorem \ref{thm:BML}, the uniqueness of parameters can be obtained. 

\begin{theorem}
\label{thm:JML}
    A binary relation $\succsim$ over menus admits a JML representation $(u, \Pi)$ if and only if it satisfies \textit{basic axioms}, \textit{completeness}, \textit{unambiguous transitivity}, and \textit{favorable mixing monotonicity}.

    Furthermore, if $\succsim$ admits another JML representation $(u', \Pi')$, then $u' = \alpha u + \beta$ for some $(\alpha, \beta ) \in \mathbb{R}_{++} \times \mathbb{R}$ and $\Pi = \Pi'$. 
\end{theorem}

Thus, the differences between BML and JML preferences lie in the axioms of completeness and transitivity.\footnote{Note that the other axioms required by either theorem are satisfied by both preferences.} 
These axioms capture how each class of preferences addresses disagreements in beliefs about states and predictions regarding future information that the members have.

\section{Comparing Rationality}

The preferences studied in the previous section violate rationality due to disagreements among members, and sets of information structures capture the degree of disagreement. 
In this section, we formally examine the relationships between rationality and the degree of disagreement by conducting comparative statics.

Consider two DMs, DM1 and DM2, with preferences $\succsim_1$ and $\succsim_2$ admitting  BML or JML representations $(u_1, \Pi_1)$ and $(u_2, \Pi_2)$, respectively. 
To focus on the degree of disagreement, we assume throughout this section that $u_1 = u_2 = u$ for some  nonconstant affine function $u:X\rightarrow \mathbb{R}$. 

\subsection{BML preferences}

A DM with a BML preference evaluates one menu as weakly better than another if all members agree, but remains silent in cases of disagreement.
Therefore, the level of decisiveness is directly related to the degree of disagreement among the members. 
To compare two binary relations in terms of decisiveness, we introduce the following definition.

\begin{definition}
    For binary relations $\succsim_1$ and $\succsim_2$ over $\mathbb{F}$,  $\succsim_1$ is \textit{more decisive} than $\succsim_2$ if  for all  $F, G\in \mathbb{F}$, $F\succsim_2 G$ implies $F\succsim_1 G$. 
\end{definition} 

\noindent
That is, $\succsim_1$ is more decisive than $\succsim_2$ 
if $\succsim_2$ is a subrelation of $\succsim_1$.  If DM2 can compare a pair of menus, then DM1 evaluates those menus in the same manner.



Another way to compare the degree of disagreement is to consider how inconsistent preferences are. Suppose that a DM with a BML preference $(u, \Pi)$ thinks  $H \not \succsim G$ and $G \not\succsim F$. 
Then there exist $\pi, \pi' \in \Pi$ such that $b^u_G (\pi) - b^u_H (\pi) > 0$ and $b^u_F (\pi') - b^u_G (\pi') > 0$. 
This means that $\pi$ and $\pi'$ are sources of evaluations $H \not \succsim G$ and $G \not\succsim F$, respectively. 
However, if $b^u_G (\pi) - b^u_F (\pi)$ and $b^u_H (\pi') - b^u_G (\pi')$ are both positive and sufficiently large, then $b^u_H (\pi) - b^u_F (\pi) > 0$ and $b^u_H (\pi') - b^u_F (\pi') > 0$, which implies that the DM does not necessarily conclude $H \not\succsim F$.
This inconsistency stems from the disagreement in information structures. If the sources of evaluations remain unchanged (i.e., $\pi = \pi'$), then $b^u_G (\pi) - b^u_H (\pi) > 0$ and $b^u_H (\pi') - b^u_G (\pi') > 0$ (or $b^u_F (\pi') - b^u_G (\pi') > 0$ and $b^u_G (\pi) - b^u_F (\pi) > 0$) cannot hold simultaneously. 



We formalize how to compare the inconsistency of the DMs’ preferences using the frequency of the inconsistent preference patterns described above.\footnote{Note that  $\succsim$ is said to satisfy \textit{negative transitivity} if for all $F',G',H' \in \mathbb{F}$, $F' \not\succsim G'$ and $G' \not\succsim H'$ imply $F' \not\succsim H'$. The preference pattern in the previous paragraph violates this property.}  
Before stating the formal definition, we introduce a notation.
For $F,G\in\mathbb{F}$, we say that $F$ \textit{strictly state-wise dominates} $G$, denoted by $F\triangleright_{D} G$, if for any $g\in G$, there exists $f\in F$ such that $f(\omega)\succ_1 g(\omega)$ for all $\omega\in\Omega$.
It should be noted that $H \triangleright_D F$ does not depend on whether the relation $\succsim_1$ or $\succsim_2$ is used to construct $\triangleright_D$ since these preferences coincide in $X$.

\begin{definition}
    For binary relations $\succsim_1$ and $\succsim_2$ over $\mathbb{F}$ such that they coincide in $X$, $\succsim_1$ is \textit{less negative-inconsistent} than $\succsim_2$ if  for all  $F, G, H\in \mathbb{F}$ such that $H \triangleright_D F$, $H \not\succsim_1 G \not\succsim_1 F$ implies $H \not\succsim_2 G \not\succsim_2 F$. 
\end{definition}

In the above definition, we consider three menus $F,G,H\in \mathbb{F}$ such that $H$ strictly state-wise dominates $F$. 
Since we study the DMs with BML preferences,  this condition implies that $H$ is strictly better than $F$. 
Therefore, the pair of evaluations $H \triangleright_D F$ and $H \not\succsim_1 G \not\succsim_1 F$ can be considered  inconsistent.
In other words, we say that $\succsim_1$ is less negative-inconsistent than $\succsim_2$ if DM2 exhibits these inconsistent preference patterns whenever DM1 does. 
We use the term ``negative-inconsistent" because we focus on the inconsistency of the negative part $\not\succsim$ of $\succsim$.

The following result shows that the degree of disagreement can be compared using the notions introduced in this section. 

\begin{theorem}
\label{thm_bml_comp}
    Let $\succsim_1$ and $\succsim_2$ be BML preferences $(u, \Pi_1)$ and $(u, \Pi_2)$, respectively. The following two statements are equivalent:
    \begin{enumerate}[(i)]
    \setlength{\itemsep}{0cm}
    \setlength{\parskip}{0cm}
        \item $\Pi_1 \subset \Pi_2$ holds. 
        \item $\succsim_1$ is more decisive than $\succsim_2$. 
    \end{enumerate}
    Furthermore, if $\Pi_1$ is not a singleton, then the following statement is also equivalent: 
    \begin{enumerate}[(i)]
    \setlength{\itemsep}{0cm}
    \setlength{\parskip}{0cm}
    \setcounter{enumi}{2}
        \item  $\succsim_1$ is less negative-inconsistent than $\succsim_2$. 
    \end{enumerate}
\end{theorem}

Notice that we omit the case where $\Pi_1$ is a singleton in the latter part of the above theorem  because,  in this case, a BML preference reduces to an SL preference and does not  exhibit  the preference patterns $H \triangleright_D F$ and $H \not\succsim_1 G \not\succsim_1 F$.

\subsection{JML preferences}

We then consider the case where DM1 and DM2 have JML preferences. 
A DM with a JML preference evaluates one menu as strictly better than another only if all members agree on that evaluation.
Consequently, severe disagreement among members prevents the DM from determining which menu is strictly better.
This observation suggests that under JML preferences, the frequency with which the DM can conclude that one menu is strictly better than another is closely linked to the degree of disagreement among the members. 
A way to compare the degree of strict decisiveness can be formalized as follows:

\begin{definition}
    For binary relations $\succsim_1$ and $\succsim_2$ over $\mathbb{F}$, we say that  $\succsim_1$ is \textit{more strict-decisive} than $\succsim_2$ if  for all  $F, G\in \mathbb{F}$, $F\succ_2 G$ implies $F\succ_1 G$. 
\end{definition} 

\noindent
That is, $\succsim_1$ is more strict-decisive than $\succsim_2$ 
if $\succ_2$ is a subrelation of $\succ_1$. 
This means that for any pair of menus, if DM2 can decide which is strictly better than the other, then DM1 makes the same decision. 

Another way to compare the degree of disagreement is to  examine the inconsistency of preferences, as in the previous subsection.
Suppose that a DM with a JML preference $(u, \Pi)$ thinks  $F \succsim G$ and $G \succsim H$. 
If the set $\Pi$ includes multiple information structures, then a consistency problem arises similarly to the BML preferences.
By $F \succsim G$ and $G \succsim H$,  there exist $\pi, \pi' \in \Pi$ such that $b^u_F (\pi) - b^u_G (\pi) \geq 0$ and $b^u_G (\pi') - b^u_H (\pi') \geq 0$. 
However, if $b^u_G (\pi) - b^u_H (\pi)$ and $b^u_F (\pi') - b^u_G (\pi')$ are both negative and sufficiently small, then $b^u_F (\pi) - b^u_H (\pi) < 0$ and $b^u_F (\pi') - b^u_H (\pi') <  0$, which implies that the DM does not necessarily conclude $F \succsim H$. 

The following definition provides a formal way to compare the degree of inconsistency using the above preference patterns.

\begin{definition}
    For binary relations $\succsim_1$ and $\succsim_2$ over $\mathbb{F}$ such that they coincide in $X$, $\succsim_1$ is \textit{less inconsistent} than $\succsim_2$ if for all  $F, G, H\in \mathbb{F}$ such that $H \triangleright_D F$, $F \succsim_1 G \succsim_1 H$ implies $F \succsim_2 G \succsim_2 H$. 
\end{definition}

In the above definition, since $H \triangleright_D F$ and the DMs have JML preferences,  $H \succ_1 F$ and $H \succ_2  F$. 
Therefore, $F \succsim_1 G \succsim_1 H$ (resp. $F \succsim_2 G \succsim_2 H$) can be seen as a violation of transitivity.

In the JML preferences, we can compare sets of information structures in terms of set inclusion using these two ways of comparison. 
The following theorem provides the formal statement.

\begin{theorem}
\label{thm_jml_comp}
    Let $\succsim_1$ and $\succsim_2$ be JML preferences $(u, \Pi_1)$ and $(u, \Pi_2)$, respectively. The following two statements are equivalent:
    \begin{enumerate}[(i)]
    \setlength{\itemsep}{0cm}
    \setlength{\parskip}{0cm}
        \item $\Pi_1 \subset \Pi_2$ holds. 
        \item $\succsim_1$ is more strict-decisive than $\succsim_2$. 
    \end{enumerate}
    Furthermore, if $\Pi_1$ is not a singleton, then the following statement is also equivalent: 
    \begin{enumerate}[(i)]
    \setlength{\itemsep}{0cm}
    \setlength{\parskip}{0cm}
    \setcounter{enumi}{2}
        \item  $\succsim_1$ is less inconsistent than $\succsim_2$. 
    \end{enumerate}
\end{theorem}

In the BML and JML preferences, disagreements among members are represented by a set of information structures. 
The results in this section show that the degree of disagreement and violations of rationality are logically related as well.

\section{Generalizations}

We considered preferences that do not satisfy either \textit{completeness} or \textit{transitivity}; so
our next question is how can we characterize preferences without them. 
In this section, we examine a general class of preferences that includes BML and JML preferences as special cases.
We find that by imposing the axioms shared by these two classes of preferences, we obtain the following class of decision criteria.

\begin{definition}
\label{def:hml}
     For a nonconstant affine function $u: X \rightarrow \mathbb{R}$ and a nonempty compact collection $\mathbf{\Pi}$ of nonempty closed convex subsets of $\Delta (\Delta (\Omega ))$, 
     a binary relation $\succsim$ over menus admits a \textbf{\textit{hierarchical multiple learning (HML) representation}} $(u,\boldsymbol{\Pi})$ if for all $F, G \in \mathbb{F}$,
    \begin{equation}
    \label{eq:HML}
        F \succsim G \iff \max_{\Pi \in \mathbf{\Pi}} \min_{\pi \in \Pi} (b_F^u (\pi) -  b_G^u (\pi)) \geq 0. 
    \end{equation}
\end{definition}

\noindent
In the context of collective decision making, a DM with an HML preference $(u,\boldsymbol{\Pi})$ can be interpreted as comparing two menus in the following way: Members are divided into several sub-groups, represented by the set $\boldsymbol{\Pi}$.
In each sub-group $\Pi\in\mathbf{\Pi}$, evaluations of the menus are made using the unanimity rule.
If some sub-group determines that one menu is better than the other, the DM adopts that judgment.


If $\mathbf{\Pi}$ is a singleton, it reduces to a BML preference. 
On the other hand, if each $\Pi \in \mathbf{\Pi}$ is a singleton, it reduces to a JML preference.
Furthermore, by setting $\boldsymbol{\Pi}=\{\alpha\Pi+(1-\alpha)\{\pi\}:\pi\in\Pi\}$, we can observe that HML preferences include an $\alpha$-maxmin type preference 
\begin{equation*}
    \alpha\min_{\pi\in\Pi}(b_F^u (\pi) -  b_G^u (\pi))+(1-\alpha)\max_{\pi\in\Pi}(b_F^u (\pi) -  b_G^u (\pi))\geq 0
\end{equation*}
as a special case.

Due to this generality, the HML preferences satisfy neither \textit{completeness} nor \textit{transitivity}. 
To characterize the HML preferences, we introduce the reflexivity axiom as a minimal requirement for the rationality.
\begin{axiom}[Reflexivity]
    For all $F\in\mathbb{F}$, $F\succsim F$.
\end{axiom}
The following theorem shows that the HML preferences are characterized by the axioms shared by the BML and JML preferences, in addition to \textit{reflexivity}.

\begin{theorem}
\label{thm:HML}
    A binary relation $\succsim$ over menus admits an HML representation if and only if it satisfies \textit{basic axioms}, \textit{completeness for lotteries}, \textit{unambiguous transitivity}, and \textit{reflexivity}.  
\end{theorem}
 
It should be noted that the order of the maximization operator and the minimization operator is inconsequential.
In the proof of Theorem \ref{thm:HML}, we construct $\mathbf{\Pi}$ by representing the upper contour set of the origin of $\mathbb{R}^{\Delta (\Omega)}$ by a union of convex cones. 
This operation can be done since this upper contour set is a cone.
Because the complement of the upper contour set  is also a cone, we can proceed a similar argument for its complement.
Consequently, we can obtain a min-of-max representation.
That is, any HML representation can be rewritten as for some a nonempty compact collection $\mathbf{\Pi}'$ of nonempty closed convex subset of $\Delta(\Delta(\Omega))$ such that for all $F, G \in \mathbb{F}$, 
\begin{equation}
\label{eq:HML_reverse}
        F \succsim G \iff \min_{\Pi' \in \mathbf{\Pi}'} \max_{\pi' \in \Pi'} (b_F^u (\pi') -  b_G^u (\pi')) \geq 0. 
\end{equation}
In general, if a preference $\succsim$ is represented as \eqref{eq:HML} and \eqref{eq:HML_reverse}, $\mathbf{\Pi}$ does not coincide with $\mathbf{\Pi}'$. 
Therefore, once a collection is fixed, the order of applying the two operators is no longer interchangeable.

\section{Rationalization Procedures}

We examined preferences that violate \textit{completeness} and/or \textit{transitivity}. 
However, when considering the real-world behavior of collective DMs, such as managers at companies or politicians, they are often not allowed to be silent, and their choices are expected to have no cycle.
That is, DMs must act based on a rational criterion that is both complete and transitive.
To address these issues, this section explores how they should establish rational criteria from the incomplete and/or intransitive preferences discussed in the previous sections.
We call these operations \textit{rationalization procedures}. 

To cover all the preferences discussed in the previous sections, we examine the rationalization procedures of the HML preferences.
This directly leads to the rationalization procedures of the BML and JML preferences as corollaries.
As in the previous sections, we denote by $\succsim$ an incomplete and/or intransitive preference over $\mathbb{F}$.
This represents the DM's first criterion, which is constructed based on members’ predictions, and the DM forms her rational criterion, denoted by $\succsim^\land$. 
We identify a class of admissible rational preferences by imposing axioms about the relationship between $\succsim$ and $\succsim^\land$. 


In addition to \textit{completeness} and \textit{transitivity}, we impose \textit{mixture continuity} on $\succsim^\land$, which is a weaker continuity axiom than \textit{continuity}  imposed in the previous sections. 

\begin{axiom}[Mixture continuity]
\label{axiom:mixcon}
    For all $F, G, H \in \mathbb{F}$, the following sets are closed: 
\begin{equation*}
        \{ \alpha \in [0,1] : \alpha F + (1 - \alpha ) G \succsim H \} ~~~ \text{and} ~~~ \{ \alpha \in [0,1] : H \succsim \alpha F + (1 - \alpha ) G \}.
\end{equation*}
\end{axiom}

Next, we introduce axioms that connect $\succsim$ to $\succsim^\land$. 
The first axiom requires that when comparing two constant acts $x$ and $y$, if the first criterion $\succsim$ states that a lottery $x$ is weakly better than a lottery $y$, then the second rational criterion should reach the same conclusion. 

\begin{axiomB}[Lottery consistency]
    For all $x,y \in X$,  if $x\succsim y$, then $ x\succsim^\land y$. 
\end{axiomB}

The second axiom deals with any pair of menus. 
Before presenting the formal definition, we introduce a binary relation derived from $\succsim$. In the real world, data about alternatives is often noisy, meaning that it may include misspecification. Such noise can influence how menus are evaluated, potentially leading to reversals in their ranking. Consequently, the observed preference patterns may not be entirely reliable. To address this issue, we focus on evaluations between two menus that are unaffected by such noise.

A ``robust’’ binary relation, in the sense described above, is defined as follows: 
Given $\succsim$, let $\stsucc$ be the binary relation over $\mathbb{F}$ such that  for all $F,G\in \mathbb{F}$, $F\stsucc G$ if and only if for all $x,y\in X$, there exists $\delta > 0$ such that for all $\varepsilon \in (0, \delta)$, 
\begin{equation*}
    (1 - \varepsilon) F +\varepsilon x \succ (1 - \varepsilon) G + \varepsilon y.
\end{equation*}
That is, we write $F \stsucc G$ when a menu $F$ is strictly better than a menu $G$ even if considering noise in observed data.
\citet{cerreia2020rational} also considered a similar binary relation in the Anscombe-Aumann framework. 

Note that for a binary relation $\succsim$  admitting an HML representation $(u, \mathbf{\Pi})$, 
\begin{align*}
    F\succ G 
    &\iff \qty[  F\succsim G~~~~ \text{and} ~~~~ G ~\cancel{\succsim} ~ F ] 
    \\
    &\iff \qty[ \max_{\Pi \in \mathbf{\Pi}} \min_{\pi \in \Pi} (b_F^u (\pi) -  b_G^u (\pi)) \geq 0 ~~~~ \text{and} ~~~~ \max_{\Pi \in \mathbf{\Pi}} \min_{\pi \in \Pi} (b_G^u (\pi) -  b_F^u (\pi)) < 0 ].
\end{align*}
Therefore, the binary relation $\stsucc$ can be rewritten as  for all $F,G\in \mathbb{F}$, 
\begin{align*}
    F\stsucc G 
    &\iff \qty[ \max_{\Pi \in \mathbf{\Pi}} \min_{\pi \in \Pi} (b_F^u (\pi) -  b_G^u (\pi)) > 0 ~~~~ \text{and} ~~~~ \max_{\Pi \in \mathbf{\Pi}} \min_{\pi \in \Pi} (b_G^u (\pi) -  b_F^u (\pi)) < 0 ]. 
\end{align*}

We then present the second axiom using $\stsucc$. 
It postulates that when comparing two menus $F$ and $G$, the second rational preference concludes that $F$ is better than $G$ if there exists $x\in X $ such that $F$ is robustly better than $x$ and $G$ is robustly worse than $x$.

\begin{axiomB}[Robustly strict consistency]
    For all $F, G \in \mathbb{F}$, if there exists $x\in X$ such that $F\stsucc x\stsucc G$, then $ F\succ^\land G$. 
\end{axiomB}

A DM with an HML preference as her first criterion can fully understand the value of constant acts ex ante since her value does not depend on states. 
Although the value of a menu changes according to information structures in general, $F\stsucc x$  provides an unambiguous lower bound for the value of $F$ using an outcome $x$.
Similarly, $x\stsucc G$ provides an unambiguous upper bound for the value of $G$.
In summary, $F\stsucc x\stsucc G$ can be interpreted as indicating that $F$ is better than $G$ regardless of information structures the DM uses to evaluate each menu.
Furthermore, this relation is robust to small noise in the observed data. 
\textit{Robustly strict consistency} postulates the consistency between the two binary relations if one menu is better than another menu in the above sense. 

By considering these two axioms, we obtain a general class of rational preferences.
A DM with such a preference evaluates each menu by the weighted sum of the max-of-min value and the min-of-max value, with the weight varying based on the menus.

\begin{theorem}
\label{thm_alpha_rat}
    Let $\succsim$ be an HML preference $(u, \mathbf{\Pi})$ and $\succsim^\land$ be a binary relation that satisfies \textit{completeness}, \textit{transitivity}, and \textit{mixture continuity}. 
    The following statements are equivalent: 
    \begin{enumerate}
        \item The pair $(\succsim, \succsim^\land)$ satisfies \textit{lottery consistency} and \textit{robustly strict consistency}. 
        \item There exists  a function $\alpha : \mathbb{F} \rightarrow [0,1]$ such that the following function $U: \mathbb{F} \rightarrow \mathbb{R}$ represents $\succsim^\land$: For all $F\in\mathbb{F}$,
        \begin{equation}
        \label{eq:alpha-HML}
            U(F) = \alpha (F) \max_{\Pi \in \mathbf{\Pi}} \min_{\pi \in \Pi} b_F^u (\pi)  + (1 - \alpha (F)) \min_{\Pi \in \mathbf{\Pi}} \max_{\pi \in \Pi} b_F^u (\pi) .
        \end{equation} 
    \end{enumerate}
\end{theorem}

\begin{remark}
    In the theorem above, we imposed \textit{mixture continuity} outside the if-and-only-if statement. 
    It is important to note that preferences represented as \eqref{eq:alpha-HML} do not necessarily satisfy \textit{mixture continuity}. However, they satisfy the following weak version of continuity: For all $F \in \mathbb{F}$ and $x, y\in X$, the sets $ \{ \alpha \in [0,1] : \alpha x + (1 - \alpha ) y \succsim F \} $ and $\{ \alpha \in [0,1] : F \succsim \alpha x + (1 - \alpha ) y \}$ are closed. 
    Examining the gap between \textit{mixture continuity} and the class of preferences represented as \eqref{eq:alpha-HML} is left for future work. 
\end{remark}

Note that it is known that the invariant bi-separable preferences over acts (i.e., weak orders that satisfy mixture continuity, state-wise monotonicity, non-triviality, and certainty independence) can be represented as the generalized $\alpha$-maxmin forms (cf. Theorem 11 of \citet{ghirardato2004differentiating}). 
In their representations, the DM with multiple priors in mind evaluates an act by the weighted sum of the expected utility levels in the best and worst scenarios, where the weights on the scenarios are act-dependent. 
These representations can be obtained in the context of rationalization of incomplete and/or intransitive preferences as well (\citet{danan2016robust,echenique2022twofold}). 

The preferences described in Theorem \ref{thm_alpha_rat} exhibit a similar structure since they examine two dual scenarios and subsequently aggregate them using a menu-dependent weighted sum.
However, they include more general considerations in each scenario. 
In the max-of-min term, the DM chooses her information structure by optimistically aggregating their sub-groups' predictions, each of which is obtained by cautiously aggregating them. 
On the other hand, in the min-of-max term, the DM chooses her information structure by cautiously aggregating their groups' predictions, each of which is obtained by optimistically aggregating them
(note that the members' predictions do not change between two scenarios since the DM uses the same collection $\mathbf{\Pi}$ of subsets in \eqref{eq:alpha-HML}; the only difference is the way of aggregating their predictions). 
After considering these two scenarios, the DM computes a weighted sum of these values.

In the generalized $\alpha$-maxmin representations under uncertainty, the weight for each act reflects how cautiously the DM evaluates that act. 
On the other hand, in \eqref{eq:alpha-HML}, we cannot interpret $\alpha (F)$ in the same manner  because whether max-of-min
or min-of-max yields a larger value  depends on the menus evaluated by the DM.

Using the theorem above, we can obtain a corollary for the BML and JML preferences. 

\begin{corollary}
    Let $\succsim$ be a BML or JML preference $(u, \Pi)$ and $\succsim^\land$ be a binary relation that satisfies \textit{completeness}, \textit{transitivity}, and \textit{mixture continuity}. 
    The following statements are equivalent: 
    \begin{enumerate}
        \item The pair $(\succsim, \succsim^\land)$ satisfies \textit{lottery consistency} and \textit{robustly strict consistency}. 
        \item There exists  a function $\alpha : \mathbb{F} \rightarrow [0,1]$ such that the following function $U: \mathbb{F} \rightarrow \mathbb{R}$ represents $\succsim^\land$: For all $F\in\mathbb{F}$,
        \begin{equation}
        \label{eq:alpha-BJ}
            U(F) = \alpha (F)  \min_{\pi \in \Pi} b_F^u (\pi)  + (1 - \alpha (F)) \max_{\pi \in \Pi} b_F^u (\pi) .
        \end{equation} 
    \end{enumerate}
\end{corollary}

This class of preferences is the direct counterpart of the generalized $\alpha$-maxmin preferences. In \eqref{eq:alpha-BJ}, the weight can be viewed as indicating how cautiously the DM evaluates each menu.


\section*{Appendix A: Notation and Mathematical Results}
\label{sec:math}

\subsection*{A.1 Mathematical preliminaries}

Let $C(\Delta(\Omega))$ be the set of real-valued continuous functions on $\Delta(\Omega)$ endowed with the supnorm. The norm dual of $C(\Delta(\Omega))$ is the set of signed Borel measures of bounded variations endowed with the weak$^\ast$ topology, denoted by  $ca (\Delta (\Omega))$ (cf. Corollary 14.15 of \citet{AB2006Math}). 
Let $ca (\Delta (\Omega))^\ast$ be the norm dual of $ca (\Delta (\Omega))$ endowed with the weak topology. The set $C(\Delta(\Omega))$ is dense in $ca (\Delta (\Omega))^\ast$ (cf. Theorem 6.24(3) of \citet{AB2006Math}). 

We denote the set of twice continuously differentiable functions on $\Delta(\Omega)$ by  $C^2 (\Delta(\Omega))$. 
It is known that $C^2 (\Delta(\Omega))$ is dense in $C (\Delta(\Omega))$ (cf. Theorem 11.3 of \citet{aliprantis1998principles}).

Let $\Phi$ be the set of convex functions on $\Delta(\Omega)$. 
For any twice continuously differentiable function $\varphi \in C^2 (\Delta(\Omega))$, there exist $\phi_1, \phi_2 \in \Phi$ such that $\varphi = \phi_1 - \phi_2$ (cf. Theorem 2.2 and  the subsequent discussion of \citet{hiriart1985generalized}).

Let $u : \Delta (X) \rightarrow \mathbb{R}$ be an affine function. 
For any $F\in \mathbb{F}$, define $\varphi_F \in \Phi$ as for all $p\in \Delta(\Omega)$, 
\begin{equation*}
    \varphi_F (p) = \max_{f\in F} \int_\Omega u(f(\omega)) p (d\omega). 
\end{equation*}
Let $\Phi_\mathbb{F} = \{ \varphi_F \in \Phi : F\in \mathbb{F} \}$. The cone generated from $\Phi_\mathbb{F}$, denoted by $\text{cone} (\Phi_\mathbb{F})$, satisfies the following property: The set $\text{cone} (\Phi_\mathbb{F}) + \mathbb{R}$ is dense in $\Phi$. 
We can prove this property by a minor modification of the arguments in \citet{de2017rationally}.

\subsection*{A.2 Variants of the separating hyperplane theorem}

We provide two variants of the separating hyperplane theorem. The first one is about separation of two points in $ca (\Delta (\Omega)) $, and the second one is about separation of two nonempty convex closed sets in $ca (\Delta (\Omega))$.\footnote{\citet{dekel2001representing} used similar techniques in a different domain from ours. For details, see Appendix C.5 of \cite{dekel2001representing}.}

\begin{proposition}
    For any two points $\pi, \pi' \in ca (\Delta (\Omega))$ such that $\pi \neq \pi'$, there exists a convex function $\varphi \in\Phi$ such that  $\langle \varphi, \pi \rangle > \langle  \varphi ,  \pi' \rangle$ or $\langle \varphi, \pi' \rangle > \langle  \varphi ,  \pi \rangle$. 
\end{proposition}

\begin{proof}
    Suppose to the contrary that for any convex  function $\varphi \in \Phi$, $\langle \varphi,  \pi   \rangle = \langle \varphi,  \pi'  \rangle $. 
    For any $\varphi' \in C^2 (\Delta (\Omega))$, there exist  $\psi_1, \psi_2 \in \Phi$ such that $\varphi' = \psi_1 - \psi_2$. 
    Then, we have $\langle \varphi',  \pi - \pi'  \rangle = \langle\psi_1,  \pi - \pi'  \rangle + \langle \psi_2,  \pi - \pi'  \rangle = 0$. 
    Since $C^2 (\Delta (\Omega))$ is dense in $C(\Delta (\Omega))$, for all $\phi \in C(\Delta (\Omega))$, $\langle \pi  , \phi \rangle = \langle \pi'  , \phi \rangle $. This means that $\pi = \pi'$, a contradiction. 
\end{proof}

\begin{proposition}
\label{prop:quasi_hyperplane}
    For all two nonempty convex closed sets $\Pi, \Pi' \subset ca (\Delta (\Omega))$ such that $\Pi \cap \Pi' = \emptyset$, there exist convex functions $\varphi, \psi \in\Phi $ such that for all $\pi \in \Pi$ and $\pi' \in \Pi$, 
    \begin{equation}
    \label{eq:quasi_separation}
        \langle \varphi, \pi \rangle > \langle \psi, \pi \rangle ~~~\text{and} ~~~ \langle \psi, \pi' \rangle > \langle \varphi, \pi' \rangle . 
    \end{equation}
\end{proposition}

\begin{proof}
    Since $C(\Delta(\Omega))$ is dense in $ca (\Delta (\Omega))^\ast$, by applying the separating hyperplane theorem, there exists $\phi \in C(\Delta (\Omega))$ such that for all $\pi \in \Pi$ and $\pi' \in \Pi$, 
    \begin{equation}
    \label{eq:sepa_1}
        \langle \phi, \pi \rangle  > 0 > \langle \phi, \pi' \rangle . 
    \end{equation}
    Furthermore, since $C^2 (\Delta (\Omega))$ is dense in $C( \Delta (\Omega) )$, we can assume $\phi \in C^2 (\Delta (\Omega))$. 
    Then there exist convex functions $\varphi, \psi \in \Phi$ such that $\phi = \varphi - \psi$. 
    By \eqref{eq:sepa_1}, we have 
    \begin{equation*}
        \langle \varphi - \psi, \pi \rangle  > 0 > \langle \varphi - \psi, \pi' \rangle, 
    \end{equation*}
    which is equivalent to \eqref{eq:quasi_separation}. 
\end{proof}

\section*{Appendix B: Preliminary Proofs}

This section provides the common part of the proofs for our characterization results in the main text.

Consider a preference $\succsim$ over menus that satisfies \textit{basic axioms}, \textit{reflexivity}, \textit{unambiguous transitivity}, and \textit{completeness for lotteries} throughout this section.
Note that preferences considered in Theorems \ref{thm:BML} and \ref{thm:JML} satisfy these axioms  since \textit{reflexivity} is implied by either \textit{dominance} or \textit{completeness}. 

\begin{claim}
\label{claim:Utrans_p4fD}
    $\succsim$ satisfies \textit{preference for flexibility} and \textit{dominance}.
\end{claim}

\begin{proof}
    We first prove that \textit{unambiguous transitivity} implies \textit{preference for flexibility}.
    Consider any $F,G\in\mathbb{F}$ such that $G\subset F$.
    Since any $g\in G$ is also in $F$,  $F\trianglerighteq_{D}G$.
    Since $G\sim G$ follows from \textit{reflexivity}, \textit{unambiguous transitivity} implies $F\succsim G$.

    Next, we prove that \textit{unambiguous transitivity} implies \textit{dominance}.
    Take any $F\in\mathbb{F}$ and $g\in\mathcal{F}$ such that $F\trianglerighteq_{D}\{g\}$.
    By the definition of $\trianglerighteq_{D}$, we have  $F\trianglerighteq_{D}F\cup\{g\}$.
    Since $F\cup\{g\}\sim F\cup\{g\}$ follows from \textit{reflexivity}, \textit{unambiguous transitivity} implies $F\succsim F\cup\{g\}$.
    By \textit{preference for flexibility}, $F\cup\{g\}\succsim F$.
    Thus, $F\sim F\cup\{g\}$.
\end{proof}

\begin{claim}
    There exists a nonconstant affine function $u: X \rightarrow \mathbb{R}$ such that for all $x, y\in X$, $x\succsim y $ if and only if $u(x) \geq u(y)$. 
\end{claim}

\begin{proof}
    By \textit{independence}, for all $x,y,z\in X$  and $\alpha \in (0,1)$,  if  $x\sim y$, then ${1 \over 2} x + {1 \over 2} z \sim {1 \over 2} y + {1 \over 2} z$.
    Therefore, by Theorem 8 of \citet{herstein1953axiomatic}, 
    there exists an affine function $u: X \rightarrow \mathbb{R}$ such that for all $x, y\in X$, $x\succsim y $ if and only if $u(x) \geq u(y)$. 

    Finally, we prove that $u$ is nonconstant. Suppose to the contrary that $u$ is constant, that is, for all $x,y\in X$, $x\sim y$. Then, for any $F, G \in \mathbb{F}$, $F\trianglerighteq_{D} G$ and $G \trianglerighteq_{D} F$ hold. By \textit{dominance}, we have $F\sim G$. This is a contradiction to \textit{nontriviality}. 
\end{proof}

Without loss of generality, we can choose a function $u$ as $0\in \text{int} ~ u(X)$. 
Let $x_0 \in X$ be such that  $u(x_0) = 0$. 
For $F\in \mathbb{F}$, define the function $\varphi_F : \Delta (\Omega) \rightarrow \mathbb{R}$ as for all $p\in  \Delta (\Omega)$, 
\begin{equation}
\label{eq:varphi_F}
    \varphi_F (p) = \max_{f\in F} \int_\Omega u(f(\omega)) p(d\omega).
\end{equation}
Recall that $\Phi_\mathbb{F}$ is the set of real-valued functions on $\Delta (\Omega)$ with the form \eqref{eq:varphi_F}. 
We define the binary relation $\succsim^\ast$ over $\Phi_\mathbb{F}$ as for all $F, G\in \mathbb{F}$, $\varphi_F \succsim^\ast \varphi_G$ if $F\succsim G$. The symmetric part and asymmetric part are denoted by $\sim^\ast$ and $\succ^\ast$, respectively. 

\begin{claim}
\label{claim:mono}
    $\succsim^\ast$ is well-defined and monotone.
\end{claim}

\begin{proof}
    We show that $\varphi_{F}=\varphi_{G}$ implies $F\sim G$.
    First, by directly applying Claim 4 of \cite{de2017rationally}, we find that if $\varphi_{F}\geq \varphi_{G}$, then for each $g\in G$, there exists $f\in \mathrm{co}F$ such that $f(\omega)\succsim g(w)$.\footnote{For each $F\in\mathbb{F}$, $\mathrm{co}F$ is the convex hull of $F$, that is, the smallest convex set that includes $F$.}

    Next, we show that if $G\subset \mathrm{co}F$, then $F\succsim G$.
    This part can be proved in a way similar to Claim 5 of \cite{de2017rationally}, except that we use \textit{unambiguous transitivity} and \textit{indifference to ex-post randomization} instead of \textit{transitivity} and the axiom related to uncertainty loving. 
    Let $G=\{g_{1},\cdots,g_{n}\}\subset \mathrm{co}F$.
    Then, for each $g_{i}$, there exist $\alpha^{i}_{1},\cdots,\alpha^{i}_{m_{i}}> 0$ summing up to 1 and $f^{i}_{1},\cdots,f^{i}_{m_{i}}\in F$ such that $g_{i}=\sum_{j=1}^{m_{i}} \alpha^i_j f^{i}_{j}$.
    Then we have, 
    \begin{equation*}
        G\subset\sum_{j=1}^{m_{1}}\cdots\sum_{j'=1}^{m_{n}}\alpha^{1}_{j}\cdots\alpha^{n}_{j'}F=\sum_{k=1}^{l}\beta_{k}F.
    \end{equation*}
    By this inclusion, we have $\sum_{k=1}^{l}\beta_{k}F\trianglerighteq_{D}G$.
    By using \textit{indifference to ex-post randomization}, $F\sim\sum_{k=1}^{l}\beta_{k}F$ holds.
    By \textit{unambiguous transitivity}, $F\succsim G$.

    By the first argument, if $\varphi_{F}\geq\varphi_{G}$, then there exists $H\subset\mathrm{co}F$ such that $H\trianglerighteq_{D}G$.
    By the second argument, if $H\subset\mathrm{co}F$, then $F\succsim H$.
    Thus, by \textit{unambiguous transitivity}, $F\succsim G$, that is,  $\succsim^\ast$ is monotone. 
    Note that for $F, G\in \mathbb{F}$,  $\varphi_F = \varphi_G$ implies $F\sim G$, that is,  $\varphi_F \sim^\ast \varphi_G$. Therefore, $\succsim^\ast$ is well-defined. 
\end{proof}

\begin{claim}
\label{claim:ind}
    For all $F, G, H\in \mathbb{F}$ and $\alpha \in (0,1)$, $\varphi_F \succsim^\ast \varphi_G$ if and only if $\alpha \varphi_F + ( 1- \alpha) \varphi_H \succsim^\ast \alpha \varphi_G + ( 1- \alpha) \varphi_H$. 
\end{claim}

\begin{proof}
    Let $F, G, H\in \mathbb{F}$ and $\alpha \in (0,1)$. 
    Suppose that $\varphi_F \succsim^\ast \varphi_G$, that is, $F\succsim G$. 
    By \textit{independence}, it is equivalent to $\alpha F + ( 1- \alpha) H \succsim \alpha G + ( 1- \alpha) H$, that is, 
    \begin{equation}
    \label{eq:ind1}
        \varphi_{\alpha F + ( 1- \alpha) H} \succsim^\ast \varphi_{\alpha G + ( 1- \alpha) H}. 
    \end{equation}
    Note that 
    \begin{align*}
        \varphi_{\alpha F + ( 1- \alpha) H} 
        &=  \max_{(f, h) \in F\times  H} \int_\Omega \qty[ \alpha u( f(\omega) ) +  (1-\alpha) u( h(\omega) ) ] p(d\omega) \\
        &= \alpha \max_{f \in F} \int_\Omega u( f(\omega) ) p(d\omega) + ( 1- \alpha ) \max_{h \in H} \int_\Omega u( h(\omega) ) p(d\omega) \\
        &= \alpha \varphi_F + ( 1- \alpha) \varphi_H
    \end{align*}
    and similarly, $\varphi_{\alpha G + ( 1- \alpha) H} = \alpha \varphi_G + ( 1- \alpha) \varphi_H$. 
    Therefore, \eqref{eq:ind1} is equivalent to $\alpha \varphi_F + ( 1- \alpha) \varphi_H \succsim^\ast \alpha \varphi_G + ( 1- \alpha) \varphi_H$. 
\end{proof}

\begin{claim}
\label{claim:homo}
    For all $F, G \in \mathbb{F}$ and $\alpha \in \mathbb{R}_{++}$ such that $\alpha \varphi_F, \alpha \varphi_G \in \Phi_\mathbb{F}$, $\varphi_F \succsim^\ast \varphi_G$ if and only if $\alpha \varphi_F \succsim^\ast \alpha \varphi_G$. 
\end{claim}

\begin{proof}
    Let $F, G \in \mathbb{F}$ and $\alpha \in (0,1)$. Suppose that $\varphi_F \succsim^\ast \varphi_G$. Then by Claim \ref{claim:ind}, it is equivalent to $\alpha \varphi_F = \alpha \varphi_F + ( 1- \alpha) \varphi_{x_0} \succsim^\ast \alpha \varphi_G + ( 1- \alpha) \varphi_{x_0} = \alpha \varphi_G$. 

     Consider the case where $\alpha >1$ such that $\alpha \varphi_F, \alpha \varphi_G \in \Phi_\mathbb{F}$. By the result of the last paragraph, $\alpha \varphi_F \succsim^\ast \alpha \varphi_G$ is equivalent to $\varphi_F = {1\over \alpha} \alpha \varphi_F \succsim^\ast {1\over \alpha} \alpha \varphi_G  =   \varphi_G$, as required.
\end{proof}

Since $\succsim^\ast$ is homogeneous in $\Phi_\mathbb{F}$ and $0 \in \text{int}~ u(X)$, we can uniquely extend it to $\text{cone} (\Phi_\mathbb{F}) + \mathbb{R}$. 
By Claim \ref{claim:ind}, it is straightforward to prove that $\succsim^\ast$ over $\text{cone} (\Phi_\mathbb{F}) + \mathbb{R}$ also satisfies the property that for all $\varphi, \varphi', \varphi'' \in \text{cone} (\Phi_\mathbb{F}) + \mathbb{R}$ and $\alpha \in (0,1)$, 
\begin{equation}
\label{eq:ind_full}
    \varphi \succsim^\ast \varphi' \iff \alpha \varphi + ( 1- \alpha) \varphi'' \succsim^\ast \alpha \varphi' + ( 1- \alpha) \varphi''. 
\end{equation}
Furthermore, by Claim \ref{claim:homo}, the following also holds:  for all $\varphi, \varphi' \in \text{cone} (\Phi_\mathbb{F}) + \mathbb{R}$ and $\alpha \in \mathbb{R}_{++}$, 
\begin{equation}
\label{eq:homo_full}
    \varphi \succsim^\ast \varphi' \iff \alpha \varphi \succsim^\ast \alpha \varphi'. 
\end{equation}
Note that by this property and the argument in the proof of Claim \ref{claim:ind}, for all $\varphi, \varphi' \in \text{cone} (\Phi_\mathbb{F}) + \mathbb{R}$ and $\alpha \in (0,1)$, we have $\alpha \varphi + (1 - \alpha )\varphi' \in \text{cone} (\Phi_\mathbb{F}) + \mathbb{R}$. 

\begin{claim}
\label{claim:imp-unatra}
    $\succsim^\ast$ over $\mathrm{cone}(\Phi_{\mathbb{F}})+\mathbb{R}$ satisfies the following property: For all $\varphi, \varphi', \varphi'' \in \mathrm{cone}(\Phi_{\mathbb{F}})+\mathbb{R}$, if (i) $\varphi \geq \varphi'$ and $\varphi' \succsim^\ast \varphi''$ or (ii) $\varphi \succsim^\ast \varphi'$ and $\varphi' \geq \varphi''$, then $\varphi \succsim^\ast \varphi''$. 
\end{claim}

\begin{proof}
    We only prove (i). 
    First, we prove that for all $F, G, H\in \mathbb{F}$, if $\varphi_F \geq \varphi_G$ and $\varphi_G \succsim^\ast \varphi_H$, then $\varphi_F \succsim^\ast \varphi_H$. 
    By Claim 4 of \citet{de2017rationally}, there exists $F^+ \in \mathbb{F}$ such that $F^+\subset \mathrm{co}F$ and $F^+ \trianglerighteq_{D} G$. Since $\varphi_G \succsim^\ast \varphi_H$ is equivalent to $G\succsim H$, \textit{unambiguous transitivity} implies that $F^+ \succsim H$. Since $F\cup F^+ \supset F^+$, we have $F\cup F^+ \trianglerighteq_{D} F^+$. By \textit{unambiguous transitivity}, $F\cup F^+ \succsim H$, which is equivalent to $\varphi_{F\cup F^+} \succsim^\ast \varphi_H$.  Note that by $\mathrm{co}F = \mathrm{co} (F\cup F^+)$, $\varphi_F = \varphi_{F\cup F^+}$ holds. 
    Since $\succsim^\ast$ on $\Phi_\mathbb{F}$ is well-defined (Claim \ref{claim:mono}), we have $\varphi_F \succsim^\ast \varphi_H$.

    It is straightforward to prove the statement  by using \eqref{eq:ind_full} and \eqref{eq:homo_full}. 
\end{proof}

Let $\widetilde{\Phi}_\mathbb{F} = \text{cone} (\Phi_\mathbb{F}) + \mathbb{R} - ( \text{cone} (\Phi_\mathbb{F}) + \mathbb{R}) = \text{cone} (\Phi_\mathbb{F}) - \text{cone} (\Phi_\mathbb{F}) + \mathbb{R}$. 
Furthermore, extend the binary relation $\succsim^\ast$ from $\text{cone} (\Phi_\mathbb{F}) + \mathbb{R}$ to $\widetilde{\Phi}_\mathbb{F}$ as follows: 
For all $\varphi, \psi \in \widetilde{\Phi}_\mathbb{F}$, $\varphi \succsim^\ast \psi$ if and only if there exists $\varphi', \varphi'', \psi', \psi'' \in  \text{cone} (\Phi_\mathbb{F}) + \mathbb{R}$ such that $\varphi = \varphi'-  \varphi''$, $\psi =  \psi' - \psi''$, and $\varphi' + \psi'' \succsim^\ast \varphi'' + \psi'$. 

\begin{claim}
     $\succsim^\ast$ on $\widetilde{\Phi}_\mathbb{F}$ is well-defined.
\end{claim}

\begin{proof} 
    Notice that for any $\varphi,\psi\in\widetilde{\Phi}_\mathbb{F}$, there exist $\varphi', \psi'\in\mathrm{cone}(\Phi_{\mathbb{F}})+\mathbb{R}$ such that $\varphi-\psi=\varphi'-\psi'$.
    Indeed, since $\varphi=\varphi_{1}-\varphi_{2}$ and $\psi=\psi_{1}-\psi_{2}$ for some $\varphi_{1},\varphi_{2},\psi_{1},\psi_{2}\in\mathrm{cone}(\Phi_{\mathbb{F}})+\mathbb{R}$, by setting $\varphi'=\varphi_{1}+\psi_{2}$ and $\psi'=\varphi_{2}+\psi_{1}$, we can explicitly construct them.
    
    We next show that for every distinct $\varphi,\psi,\varphi',\psi'\in\mathrm{cone}(\Phi_{\mathbb{F}})+\mathbb{R}$, if $\varphi\succsim^\ast\psi$ and $\varphi-\psi=\varphi'-\psi'$, then $\varphi'\succsim^\ast\psi'$.
    Since $\varphi=\varphi'-\psi'+\psi$, $\varphi\succsim^\ast\psi$ if and only if $\varphi'-\psi'+\psi\succsim^\ast\psi$.
    By \eqref{eq:ind_full}, this is equivalent to $\frac{1}{2}(\varphi'-\psi'+\psi)+\frac{1}{2}\psi'\succsim^\ast\frac{1}{2}\psi+\frac{1}{2}\psi'$, which can be rewritten as  $\frac{1}{2}\varphi'+\frac{1}{2}\psi\succsim^\ast\frac{1}{2}\psi+\frac{1}{2}\psi'$.
    Again, by \eqref{eq:ind_full}, this is equivalent to $\varphi'\succsim^\ast\psi'$.
\end{proof}

Let $K =  \{ \phi \in  \widetilde{\Phi}_\mathbb{F}  : \phi \succsim^\ast 0 \}$. 
Then, the following result holds. 

\begin{claim}
\label{claim:K_closed}
    $K$ is a nonempty closed cone including $\{ \varphi \in  \widetilde{\Phi}_\mathbb{F} : \varphi \geq 0 \}$. 
\end{claim}

\begin{proof}
    By $0\in \text{int} ~ u(X)$ and \eqref{eq:homo_full}, it is straightforward to prove that $K$  is a nonempty cone.
    
    The fact that $K$ includes $\{ \phi \in  \widetilde{\Phi}_\mathbb{F} : \text{$\phi (p) \geq 0$ for all $p\in \Delta(\Omega)$} \}$ follows from Claim \ref{claim:mono}. 
    Indeed, for $\varphi \in \widetilde{\Phi}_\mathbb{F}$ such that $\varphi \geq 0$, take $\varphi_1, \varphi_2 \in \mathrm{cone}(\Phi_{\mathbb{F}})+\mathbb{R}$ as $\varphi = \varphi_1 - \varphi_2$. By $\varphi \geq 0$, we have $\varphi_1 \geq \varphi_2$. By Claim \ref{claim:mono}, $\varphi_1 \succsim^\ast \varphi_2$.  The definition of $\succsim^\ast$ on $\mathrm{cone}(\Phi_{\mathbb{F}})+\mathbb{R}$ implies $\varphi \succsim^\ast 0$. 
    
    We then prove that for all $\varphi,\psi\in\widetilde{\Phi}_\mathbb{F}$, the set $\{\alpha\in[0,1]:\alpha\varphi+(1-\alpha)\psi\succsim^\ast0\}$ is closed.
    Let $\varphi',\varphi'',\psi',\psi'' \in \mathrm{cone}(\Phi_{\mathbb{F}})+\mathbb{R}$ with $\varphi=\varphi'-\varphi''$ and $\psi=\psi'-\psi''$.
    Then, for all $\alpha\in[0,1]$, $\alpha\varphi+(1-\alpha)\psi\succsim^\ast0$ is equivalent to $\alpha(\varphi'-\varphi'')+(1-\alpha)(\psi'-\psi'')\succsim^\ast0$.
    By the definition of $\succsim^\ast$, this is equivalent to $\alpha\varphi'+(1-\alpha)\psi'\succsim^\ast\alpha\varphi''+(1-\alpha)\psi''$.
    By \textit{continuity}, the set of $\{\alpha\in[0,1]:\alpha\varphi'+(1-\alpha)\psi'\succsim^\ast\alpha\varphi''+(1-\alpha)\psi''\}$ is closed.
    Thus, the set $\{\alpha\in[0,1]:\alpha\varphi+(1-\alpha)\psi\succsim^\ast0\}$ is closed.

    Next, we show that $K$ is closed.
    Let $\{\varphi^n\}_{n\in\mathbb{N}}$ be a sequence in $K$ such that $\varphi^{n}\rightarrow\varphi\in\widetilde{\Phi}_\mathbb{F}$.
    Let $M=\max_{p\in\Delta(\Omega)}\varphi(p)$.
    Then, for all $\varepsilon>0$, there exists $n\in\mathbb{N}$ such that 
    \begin{equation*}
        \varphi^{n}\leq\varphi+\varepsilon\mathbf{1}_{\Delta(\Omega)}\leq\varphi+\varepsilon\{(M+1)\mathbf{1}_{\Delta(\Omega)}-\varphi\}.
    \end{equation*}
    Therefore, for all $\varepsilon\in(0,1)$, there exists $n\in\mathbb{N}$ such that 
    \begin{equation*}
        \varepsilon(M+1)\mathbf{1}_{\Delta(\Omega)}+(1-\varepsilon)\varphi\geq\varphi^n\succsim^\ast0.
    \end{equation*}
    Then, by \textit{unambiguous transitivity}, for all $\varepsilon\in(0,1)$, $\varepsilon(M+1)\mathbf{1}_{\Delta(\Omega)}+(1-\varepsilon)\varphi\succsim^\ast0$.
    (Indeed, for any $\psi, \psi' \in\widetilde{\Phi}_\mathbb{F}$ with $\psi\geq\psi' \succsim^\ast0$, take $\psi_{1},\psi_{2},\psi'_{1},\psi'_{2}\in\mathrm{cone}(\Phi_{\mathbb{F}})+\mathbb{R}$ be such that $\psi=\psi_{1}-\psi_{2}$ and $\psi'=\psi'_{1}-\psi'_{2}$.
    By construction, $\psi_{1}+\psi'_{2}\geq\psi_{2}+\psi'_{1}$.
    Since $\psi'\succsim^\ast0$, $\psi'_{1}\succsim^\ast\psi'_{2}$.
    By \eqref{eq:ind_full}, ${1\over2}\psi_{2}+{1\over2}\psi'_{1}\succsim^\ast {1\over2} \psi_{2}+{1\over2}\psi'_{2}$.
    By Claim \ref{claim:imp-unatra}, ${1\over2}\psi_{1}+{1\over2}\psi'_{2}\succsim^\ast{1\over2}\psi_{2}+{1\over2}\psi'_{2}$.
    By \eqref{eq:ind_full}, $\psi_{1}\succsim^\ast\psi_{2}$.
    By construction, this is equivalent to $\psi\succsim^\ast0$.) 
    Thus, by the result of the last paragraph, $\varphi\succsim^\ast0$.
\end{proof}

\section*{Appendix C: Proofs of the Results in the Main Text}

\subsection*{C.1 Proof of Theorem \ref{thm:BML}}

This section proves the only-if part of Theorem \ref{thm:BML}. We omit a proof of the if part since it is straightforward to verify. 
Let $\succsim$ be a binary relation that satisfies \textit{ basic axioms}, \textit{completeness for lotteries}, \textit{transitivity}, \textit{preference for flexibility}, and \textit{dominance}.  
First, to apply the results in Appendix B, we verify that $\succsim$ satisfies \textit{unambiguous transitivity}. 
Since $\succsim$ satisfies \textit{transitivity}, it is sufficient to show that for all $F, G\in \mathbb{F}$, if $F\trianglerighteq_{D} G$, then $F\succsim G$. 
We can prove this in the same way as a proof of Claim 
1 in \citet{de2017rationally} using \textit{transitivity}, \textit{preference for flexibility} and \textit{dominance}. 
Therefore, we can use all of the results in Appendix B.

\begin{claim}
\label{claim:K_convex}
    $K$ is a convex set. 
\end{claim}

\begin{proof}
    To prove that $K$ is a convex set, let $\varphi, \psi \in K$ and $\alpha \in (0,1)$. 
    Then,  there exist $\varphi_1, \varphi_2, \psi_1, \psi_2 \in \mathrm{cone}(\Phi_{\mathbb{F}})+\mathbb{R}$ such that $\varphi = \varphi_1 -  \varphi_2$, $\psi = \psi_1 -  \psi_2$,   $\varphi_1 \succsim^\ast \varphi_2$, and $\psi_1 \succsim^\ast \psi_2$. 
     By $\varphi_1 \succsim^\ast \varphi_2$ and \eqref{eq:ind_full}, 
    \begin{equation*}
        \alpha \varphi_1 + (1 - \alpha) \psi_1 \succsim^\ast \alpha \varphi_2 + (1 - \alpha) \psi_1
    \end{equation*}
    By $\psi_1 \succsim^\ast \psi_2$ and \eqref{eq:ind_full}, 
    \begin{equation*}
        \alpha \psi_1 + (1 - \alpha) \varphi_2 \succsim^\ast \alpha \psi_2 + (1 - \alpha) \varphi_2
    \end{equation*}
    Since Axiom \ref{axiom:trans} implies transitivity of $\succsim^\ast$ on $\mathrm{cone}(\Phi_{\mathbb{F}})+\mathbb{R}$, we have $\alpha \varphi_1 + (1 - \alpha) \psi_1 \succsim^\ast \alpha \varphi_2 + (1 - \alpha) \psi_2$. 
    Therefore,  $\alpha \varphi + (1 - \alpha ) \psi \in K$, that is, $K$ is a convex set.
\end{proof}

\begin{claim}
\label{claim:bewley}
     There exists a nonempty closed convex set $\Pi \subset \Delta(\Delta (\Omega))$ such that for all $\varphi \in \widetilde{\Phi}_\mathbb{F}$, $\varphi \in K$ if and only if $\langle \varphi, \pi \rangle \geq 0$ for all $\pi \in \Pi$. 
\end{claim}

\begin{proof}
    Define the set $\Pi \subset \Delta(\Delta (\Omega))$ as 
    \begin{equation*}
        \{ \pi\in \Delta(\Delta (\Omega)) : \langle \varphi, \pi \rangle \geq 0 ~ \text{for all} ~ \varphi \in K \}. 
    \end{equation*}
    First, we prove that $\Pi$ is a nonempty set. Since $K$ is a convex cone including $\{ \varphi \in  \widetilde{\Phi}_\mathbb{F} : \varphi \geq 0\}$ (cf. Claims \ref{claim:K_closed} and \ref{claim:K_convex}), by applying the separating hyperplane theorem (cf. Theorem 5.61  \citet{AB2006Math}) to the sets $\{ 0 \}$ and $K$, there exists $\pi \in ca(\Delta(\Omega))$ such that for all $\varphi\in K$, $ \langle \varphi, \pi \rangle \geq 0$.

    We prove that for all Borel sets $A$, $\pi (A) \geq 0$. 
    Let 
    \begin{equation*}
        ca(\Delta(\Omega))^+ = \{ \pi' \in  ca(\Delta(\Omega)) : \text{for all Borel set $A$, } \pi'(A) \geq 0 \}. 
    \end{equation*}
    Suppose to the contrary that $\pi \notin ca(\Delta(\Omega))^+$. Since $ca(\Delta(\Omega))^+$ is a convex set, by Proposition \ref{prop:quasi_hyperplane}, there exist  $\varphi, \psi \in \Phi$ such that for all $\pi' \in ca(\Delta(\Omega))^+ $, 
    \begin{equation*}
        \langle \varphi, \pi \rangle > \langle \psi, \pi \rangle ~~~\text{and} ~~~ \langle \psi, \pi' \rangle > \langle \varphi, \pi' \rangle . 
    \end{equation*}
    By the density of $\text{cone} (\Phi_\mathbb{F}) +\mathbb{R}$ in $\Phi$, we can assume $\varphi, \psi \in \Phi_\mathbb{F}$. 
    That is, there exist menus $F, G\in \mathbb{F}$ such that for all $\pi' \in ca(\Delta(\Omega))^+ $, 
    \begin{equation*}
        \langle \varphi_F, \pi \rangle > \langle \varphi_G, \pi \rangle ~~~\text{and} ~~~ \langle \varphi_G, \pi' \rangle > \langle \varphi_F, \pi' \rangle . 
    \end{equation*}
    The first inequality means that $\varphi_G - \varphi_F \notin K$. 
    Notice that for all $p\in\Delta(\Omega)$, the indicator function $\mathbf{1}_{\{p\}}\in ca(\Delta(\Omega))^{+}$.
    Thus, the second inequality implies that for all $p\in \Delta (\Omega)$, $\varphi_G (p) > \varphi_F (p) $, that is, $\varphi_G > \varphi_F $. 
    By Claim \ref{claim:K_closed}, we have $G\succsim F$, that is, $\varphi_G - \varphi_F \in K$. This is a contradiction. 

    Since $\pi$ is nonzero, we can assume that $\pi \in \Delta (\Delta (\Omega))$. Therefore, $\Pi$ is a nonempty set.

    Then, we prove that $\Pi$ is a closed, convex set. 
    Let $\{ \pi^n \}_{n\in \mathbb{N}} \subset \Pi$ be a sequence converging to $\tilde{\pi} \in \Delta(\Delta (\Omega))$. 
    Since $ca (\Delta (\Omega))$ is endowed with the weak$^\ast$ topology, $\pi^n \rightarrow \tilde{\pi}$ if and only if $\langle \phi,  \pi^n \rangle \rightarrow \langle \phi,  \tilde{\pi} \rangle$ in $\mathbb{R}$ for each $\phi \in C(\Delta(\Omega))$ (cf. p.212 of \citet{AB2006Math}). Therefore, for all $\varphi \in K$,  if $\langle \varphi,  \pi^n \rangle \geq 0$ for each $n\in \mathbb{N}$, then $\langle \varphi,  \tilde{\pi} \rangle \geq 0$, that is, $\tilde{\pi} \in \Pi$. 
 
    To prove the convexity, let $\pi, \pi' \in \Pi$ and $\alpha \in (0,1)$. By the definition of $\Pi$, for all $\varphi \in K$, $\langle \varphi, \pi \rangle \geq 0$ and $\langle \varphi, \pi' \rangle \geq 0$. Thus, we have 
    \begin{equation*}
        \langle \varphi, \alpha \pi + (1 - \alpha) \pi' \rangle 
        = 
        \alpha \langle \varphi,  \pi \rangle +   (1 - \alpha) \langle \varphi,  \pi' \rangle 
        \geq 0. 
    \end{equation*}
    This means $\alpha \pi + (1 - \alpha) \pi' \in \Pi$.

    Then by the definition of $\Pi$, for all $\varphi \in \widetilde{\Phi}_\mathbb{F}$, if $\varphi \in K$, then $\langle\varphi, \pi \rangle \geq 0$ for all $\pi \in \Pi$. 
    To prove the converse, suppose to the contrary that there exists $\varphi\notin K$ such that   $\langle\varphi, \pi \rangle \geq 0$ for all $\pi \in \Pi$.
    Then, by the separating hyperplane theorem (cf. Theorem 5.79 of 
    \citet{AB2006Math}), there exists $\pi' \in ca (\Delta (\Omega))$ such that for all $\psi \in K$, 
    \begin{equation}
    \label{eq:iff_separation}
        \langle \psi, \pi' \rangle \geq 0 > \langle \varphi, \pi' \rangle. 
    \end{equation}
    By the same argument of the second paragraph in this proof, we can prove that $\pi' \in \Delta (\Delta (\Omega))$. 
    By the first inequality of \eqref{eq:iff_separation}, $\pi' \in \Pi$. 
    Then, the second inequality contradicts the assumption that $\langle\varphi, \pi \rangle \geq 0$ for all $\pi \in \Pi$. 
\end{proof}

By Claim \ref{claim:bewley}, for all $F, G\in \mathbb{F}$, 
\begin{align*}
    & F\succsim G \\
    &\iff \varphi_F \succsim^\ast \varphi_G \\
    &\iff \min_{\pi \in \Pi} \langle \varphi_F - \varphi_G, \pi \rangle \geq 0 \\
    &\iff \min_{\pi \in \Pi} \int_{\Delta (\Omega)} \qty[ \max_{f\in F} \qty( \int_\Omega u(f(\omega)) p (d\omega) ) - \max_{g\in G} \qty( \int_\Omega u(g(\omega)) p (d\omega) ) ] \pi (dp) \geq 0, 
\end{align*}
that is,  $\succsim$ is a BML preference. 

Finally, we prove the uniqueness of the parameters. 
Suppose that $\succsim$ admits BML representations $(u_1, \Pi_1)$ and $(u_2, \Pi_2)$. 
It is straightforward to prove that for some $\alpha > 0$ and $\beta \in\mathbb{R}$, $u_1 = \alpha  u_2 + \beta$. 

It is sufficient to prove that $\Pi_1 = \Pi_2$. Suppose to the contrary that $\Pi_1 \neq \Pi_2$. 
Without loss of generality, we assume that there exists $\pi_1 \in \Pi_1$ such that $\pi_1 \notin \Pi_2$.  
By Proposition \ref{prop:quasi_hyperplane}, there exist convex functions $\varphi, \psi \in \Phi$ such that for all $\pi_2 \in \Pi_2$, 
\begin{equation*}
    \langle \varphi, \pi_1 \rangle > \langle \psi, \pi_1 \rangle ~~~\text{and} ~~~ \langle \psi, \pi_2 \rangle > \langle \varphi, \pi_2 \rangle . 
\end{equation*}
Since $\text{cone}(\Phi_\mathbb{F}) + \mathbb{R}$ is dense in $\Phi$, we can assume $\varphi, \psi \in \Phi_\mathbb{F}$. 
Then, there exist $F, G$ such that for all  $\pi_2 \in \Pi_2$,
    \begin{equation*}
        \langle \varphi_F, \pi_1 \rangle > \langle \varphi_G, \pi_1 \rangle ~~~\text{and} ~~~ \langle \varphi_G, \pi_2 \rangle > \langle \varphi_F, \pi_2 \rangle , 
    \end{equation*}
that is, 
    \begin{equation*}
        \langle \varphi_F, \pi_1 \rangle > \langle \varphi_G, \pi_1 \rangle ~~~\text{and} ~~~ \langle 
        \alpha \varphi_G + \beta, \pi_2 \rangle > \langle  \alpha \varphi_F + \beta, \pi_2 \rangle. 
    \end{equation*}
This implies that  under the BML representation $(u_1, \Pi_1)$, $G\succ F$ does not hold,  but under the BML representation $(u_2, \Pi_2)$, $G\succ F$. This is a contradiction. 

\subsection*{C.2 Proof of Theorem \ref{thm:JML}}

This section proves the only-if part of Theorem \ref{thm:JML}. We omit a proof of the if part since it is straightforward to verify. 
Let $\succsim$ be a binary relation that satisfies \textit{basic axioms}, \textit{completeness}, \textit{unambiguous transitivity}, and \textit{favorable mixing monotonicity}. 
Note that we can use all of the results in Appendix B. 

\begin{claim}
$K^{c}=\widetilde{\Phi}_{\mathbb{F}}\setminus K$ is convex.
\end{claim}

\begin{proof}
    By \textit{completeness}, it is straightforward to prove that $\succsim^\ast$ on $\widetilde{\Phi}_{\mathbb{F}}$ is also complete. 
    Therefore, $K^c$ can be written as $K^{c}=\{\varphi\in\widetilde{\Phi}_{\mathbb{F}}:0 \succ^\ast \varphi \}$.
    For any $\varphi,\psi\in K^{c}$, let $\varphi',\varphi'',\psi',\psi''\in\mathrm{cone}(\Phi_{\mathbb{F}})+\mathbb{R}$ with $\varphi=\varphi'-\varphi''$ and $\psi=\psi'-\psi''$.
    Then, $\varphi''\succ^\ast\varphi'$ and $\psi''\succ^\ast\psi'$.
    By \textit{favorable mixing monotonicity}, for all $\alpha\in[0,1]$, $\alpha\varphi''+(1-\alpha)\psi''\succ^\ast\alpha\varphi'+(1-\alpha)\psi'$.\footnote{
    More precisely, we use \eqref{eq:homo_full} as well since $\varphi', \varphi'', \psi'$ and $ \psi''$ are in not $\Phi_\mathbb{F}$ but $\mathrm{cone}(\Phi_{\mathbb{F}})+\mathbb{R}$. 
    }
    This is equivalent to $0 \succ^\ast \alpha(\varphi'-\varphi'')+(1-\alpha)(\psi'-\psi'')$.
    Thus, by construction, $0\succ^\ast \alpha\varphi+(1-\alpha)\psi$.
\end{proof}

\begin{claim}
    $K^{c}$ includes the set $\{\varphi\in\widetilde{\Phi}_\mathbb{F}:\varphi<0\}$.
\end{claim}

\begin{proof}
    For all $\varphi\in\widetilde{\Phi}_\mathbb{F}$ with $\varphi<0$, there exist $\varphi_1,\varphi_2\in\mathrm{cone}(\Phi_{\mathbb{F}})+\mathbb{R}$ such that $\varphi=\varphi_1-\varphi_2$.
    We prove $\varphi_2\succ^\ast\varphi_1$. 
    By $\varphi<0$, $\varphi_1<\varphi_2$ holds.
    Since $\varphi$ is continuous, $\varphi_1-\varphi_2<0$ and $\Delta(\Omega)$ is compact, there exists $\underline{m}:= \min_{p} [ \varphi_2(p)-\varphi_1(p)] >0$.
    Then, $\varphi_1<\varphi_1+\underline{m} \mathbf{1}_{\Delta(\Omega)} \leq\varphi_2$.
    Since $\succsim^\ast$ on $\mathrm{cone}(\Phi_{\mathbb{F}})+\mathbb{R}$ is monotone (cf. Claim \ref{claim:mono} and the definition of $\succsim^\ast$ on $\mathrm{cone}(\Phi_{\mathbb{F}})+\mathbb{R}$), we have $\varphi_2\succsim^\ast\varphi_1$. 
    
    Suppose to the contrary that $\varphi_2\sim^\ast\varphi_1$.
    Then, by Claim \ref{claim:imp-unatra}, $\varphi_1 \succsim^\ast \varphi_1 +\underline{m} \mathbf{1}_{\Delta(\Omega)} $. 
    Since $\succsim^\ast$ on $\mathrm{cone}(\Phi_{\mathbb{F}})+\mathbb{R}$ is monotone, we have  $\varphi_1 +\underline{m} \mathbf{1}_{\Delta(\Omega)}  \succsim^\ast \varphi_1$, which implies that $\varphi_1+\underline{m} \mathbf{1}_{\Delta(\Omega)} \sim^\ast\varphi_1$.
    By \eqref{eq:ind_full} and \eqref{eq:homo_full}, this relation is equivalent to $\underline{m} \mathbf{1}_{\Delta(\Omega) } \sim^\ast0$.
    Since $\underline{m} >0$, this implies that there exist $x,y\in X$ and $\alpha\in\mathbb{R}_{++}$ such that $\varphi_{x}>\varphi_{y}$ (i.e., $x\succ y$) and $\alpha(\varphi_{x}-\varphi_{y})\sim^\ast0$.
    By \eqref{eq:ind_full} and \eqref{eq:homo_full}, this is equivalent to $\varphi_{x}\sim^\ast\varphi_{y}$, that is,  $x\sim y$. This is  a contradiction.
    Thus, $\varphi_2\succ^\ast\varphi_1$ holds.
\end{proof}

\begin{claim}
\label{claim:JMLmain}
    There exists a nonempty convex closed set $\Pi\subset\Delta(\Delta(\Omega))$ such that for all $\varphi \in \widetilde{\Phi}_\mathbb{F}$, $\varphi\succsim^\ast0$ if and only if there exists $\pi\in\Pi$ such that $\langle\varphi,\pi\rangle \geq 0$.
\end{claim}

\begin{proof}
    Take any $\varphi\in K$.
    Let 
    \begin{equation*}
        \Pi_\varphi = \{ \pi_\varphi \in ca(\Delta(\Omega)): \langle\varphi,\pi_{\varphi}\rangle\geq 0 ~ \text{and} ~ \langle\psi,\pi_{\varphi}\rangle<0 ~ \text{for all} ~ \psi\in K^{c}\}. 
    \end{equation*}
    Since $K^{c}$ is convex and open, $\Pi_\varphi$ is nonempty (cf. Lemma 5.66 of \citet{AB2006Math}).
    Let $\Pi:=\bigcup_{\varphi\in K}\Pi_\varphi$.
    Since $K^{c}$ is the complement of $K$, $K$ can be written as 
    \begin{equation*}
        K = \bigcup_{\pi \in \Pi} \{ \varphi' \in \widetilde{\Phi}_\mathbb{F} : \langle \varphi', \pi \rangle \geq 0 \}.
    \end{equation*}
    Indeed, for any $\varphi\in K$, there exists $\pi\in\Pi_{\varphi}\subset\Pi$ such that $\langle\varphi,\pi\rangle\geq 0$.
    To prove the converse, take any $\varphi\in\widetilde{\Phi}_\mathbb{F}$ with $\langle\varphi,\pi\rangle\geq 0$ for some $\pi\in\Pi$. 
    By the definition of $\Pi$, $\langle\psi,\pi'\rangle<0$ for all $\psi\in K^{c}$ and $\pi'\in\Pi$. If $\varphi\notin K$, then this is a contradiction to $\langle\varphi,\pi\rangle\geq 0$. 
    Thus, $\varphi\in K$ if and only if there exists $\pi\in\Pi$ such that $\langle\varphi,\pi\rangle\geq 0$.
    
    
    Take any $\pi\in\Pi$.
    We next show that for any Borel set $A\subset\Omega$, $\pi(A)\geq 0$.
    That is, $\Pi\subset ca(\Delta(\Omega))^{+}$.
    Suppose to the contrary that there exists $\pi\in\Pi$ such that $\pi\notin ca(\Delta(\Omega))^{+}$.
    Since $ca(\Delta(\Omega))^{+}$ is a convex set, by Proposition \ref{prop:quasi_hyperplane}, there exist $\varphi,\psi\in\Phi$ such that for all $\pi'\in ca(\Delta(\Omega))^{+}$, 
    \begin{equation*}
        \langle\varphi,\pi\rangle>\langle\psi,\pi\rangle ~~~\text{and} ~~~ \langle\varphi,\pi'\rangle<\langle\psi,\pi'\rangle.
    \end{equation*}
    By the density of $\mathrm{cone}(\Phi_{\mathbb{F}})+\mathbb{R}$ in $\Phi$, we can assume that $\varphi,\psi\in\Phi_{\mathbb{F}}$.
    Then, there exist $F,G\in \mathbb{F}$ such that for all $\pi'\in ca(\Delta(\Omega))^{+}$,
    \begin{equation*}
        \langle\varphi_{F},\pi\rangle>\langle\varphi_{G},\pi\rangle ~~~\text{and} ~~~ \langle\varphi_{F},\pi'\rangle<\langle\varphi_{G},\pi'\rangle.
    \end{equation*}
    The first inequality implies that $\varphi_{F}-\varphi_{G}\notin K^{c}$.
    Notice that for all $p\in\Delta(\Omega)$, the indicator function $\mathbf{1}_{\{p\}}\in ca(\Delta(\Omega))^{+}$.
    Thus, the second inequality implies that for all $p\in\Delta(\Omega)$, $\varphi_{F}(p)-\varphi_{G}(p)<0$.
    Since $K^{c}$ contains the negative orthant, this implies that $\varphi_{F}-\varphi_{G}\in K^{c}$, a contradiction.
    Since $\Pi\subset ca(\Delta(\Omega))^{+}$, we can 
    set $\Pi\subset\Delta(\Delta(\Omega))$.

    We next show that $\Pi$ is closed and convex.
    Let $\{\pi^{n}\}_{n\in\mathbb{N}}\subset\Pi$ be such that $\pi^{n}\rightarrow\tilde{\pi}\in\Delta(\Delta(\Omega))$.
    Take any $\varphi\in K$ such that $\varphi\geq 0$.
    Then, for any $n\in\mathbb{N}$, $\langle\varphi,\pi^{n}\rangle\geq 0$.
    Since $ca(\Delta(\Omega))$ is endowed with the weak* topology, $\pi^{n}\rightarrow\tilde{\pi}$ implies $\langle\varphi,\pi^{n}\rangle\rightarrow\langle\varphi,\tilde{\pi}\rangle$.
    Thus, $\langle\varphi,\tilde{\pi}\rangle\geq 0$.
    Moreover, since $\langle\varphi',\pi^{n}\rangle<0$ for all $n\in\mathbb{N}$ and $\varphi'\in K^{c}$, by the same argument, $\langle\varphi',\tilde{\pi}\rangle\leq 0$ for all $\varphi'\in K^{c}$.
    Since $K^{c}$ is open, Lemma 5.66 of \cite{AB2006Math} implies that $\langle\varphi',\tilde{\pi}\rangle <0$ for all $\varphi'\in K^{c}$.
    Thus, we have $\tilde{\pi}\in\Pi$, and hence, $\Pi$ is closed.

    To prove that $\Pi$ is convex, take any $\pi,\pi'\in\Pi$ and $\alpha\in(0,1)$.
    Since $\pi,\pi'\in\Delta(\Delta(\Omega))$, for any $\varphi\in K$ such that $\varphi\geq 0$, we have $\langle\varphi,\alpha\pi+(1-\alpha)\pi'\rangle\geq 0$.
    Moreover, for any $\varphi'\in K^{c}$, $\langle\varphi',\alpha\pi+(1-\alpha)\pi'\rangle=\alpha\langle\varphi',\pi\rangle+(1-\alpha)\langle\varphi',\pi'\rangle<0$.
    Thus, $\alpha\pi+(1-\alpha)\pi'\in\Pi_{\varphi}\subset\Pi$, which means that  $\Pi$ is convex.
\end{proof}

By Claim \ref{claim:JMLmain}, for all $F, G\in \mathbb{F}$, 
\begin{align*}
    & F\succsim G \\
    &\iff \varphi_F \succsim^\ast \varphi_G \\
    &\iff \max_{\pi \in \Pi} \langle \varphi_F - \varphi_G, \pi \rangle \geq 0 \\
    &\iff \max_{\pi \in \Pi} \int_{\Delta (\Omega)} \qty[ \max_{f\in F} \qty( \int_\Omega u(f(\omega)) p (d\omega) ) - \max_{g\in G} \qty( \int_\Omega u(g(\omega)) p (d\omega) ) ] \pi (dp) \geq 0, 
\end{align*}
that is,  $\succsim$ is a JML preference. 

We can prove the uniqueness of the parameters in a similar way to Theorem \ref{thm:BML}.

\subsection*{C.3 Proof of Theorem \ref{thm_bml_comp}}

Let $\succsim_1$ and $\succsim_2$ be BML preferences $(u, \Pi_1)$ and $(u, \Pi_2)$, respectively.
We first prove that (i) implies (ii).
Suppose $\Pi_1 \subset \Pi_2$ and let $F,G\in \mathbb{F}$ such that $F\succsim_2 G$. 
Then, by the definition of the BML preferences, 
$\min_{\pi \in \Pi_2} ( b^u_F (\pi)  - b^u_G (\pi))  \geq 0$. 
Since $\Pi_1 \subset \Pi_2$, we have 
\begin{equation*}
    \min_{\pi \in \Pi_1} ( b^u_F (\pi)  - b^u_G (\pi))   \geq \min_{\pi \in \Pi_2}( b^u_F (\pi)  - b^u_G (\pi)). 
\end{equation*}
Therefore, $\min_{\pi \in \Pi_1} ( b^u_F (\pi)  - b^u_G (\pi))   \geq  0$, which means $F\succsim_1 G$. 

We then prove that (ii) implies (i). Suppose to the contrary that (ii) holds but there exists $\pi^\ast \in \Pi_1$ such that $\pi^\ast \notin \Pi_2$. By Proposition \ref{prop:quasi_hyperplane} and the density of $\mathrm{cone}(\Phi_{\mathbb{F}})+\mathbb{R}$ in $\Phi$, there exist $F, G\in \mathbb{F}$ such that for all $\pi \in \Pi_2$, 
\begin{equation*}
         \langle \varphi_F, \pi\rangle > \langle \varphi_G, \pi \rangle ~~~\text{and} ~~~ \langle \varphi_G, \pi^\ast \rangle > \langle \varphi_F, \pi^\ast \rangle . 
\end{equation*}
Thus, $F\succsim_2 G$ and $F\not\succsim_1 G$. This is a contradiction to (ii).

\vspace{5mm}

Next, we show the latter part. Suppose that $\Pi$ is not a singleton. 
We prove that (i) implies (iii). 
Let $F, G, H\in \mathbb{F}$ such that  $H \triangleright_D F$ and $H \not\succsim_1 G \not\succsim_1 F$. 
Then, since $\Pi_1 \subset \Pi_2$ and $H \not\succsim_1 G \not\succsim_1 F$, 
    \begin{equation*}
        \min_{\pi \in \Pi_2} (b^u_G (\pi)  - b^u_F (\pi)) 
        \leq \min_{\pi \in \Pi_1} (b^u_G (\pi) - b^u_F (\pi)) < 0. 
    \end{equation*}
    and 
    \begin{equation*}
        \min_{\pi \in \Pi_2} ( b^u_H (\pi) -  b^u_G (\pi)) 
        \leq \min_{\pi \in \Pi_1} ( b^u_H (\pi) -  b^u_G (\pi)) < 0. 
    \end{equation*}
Therefore, we have $H \not\succsim_2 G \not\succsim_2 F$. 

We then prove that (iii) implies (i). 
Suppose to the contrary that $\Pi_1 \not\subset \Pi_2$. 
Then, there exist $\pi^\ast, \pi^{\ast\ast} \in \Pi_1$ such that $\pi^\ast \notin \Pi_2$ and $\pi^{\ast\ast} \notin \text{conv} (\Pi_2 \cup \{ \pi^\ast \})$. (Indeed, by $\Pi_1 \not\subset \Pi_2$, there exists $\pi' \in \Pi_1 \backslash \Pi_2$. If there exists $\pi'' $ such that $\pi'' \notin \text{conv} (\Pi_2 \cup \{ \pi' \})$, then define $\pi^\ast$ and $\pi^{\ast\ast}$ as $\pi^\ast = \pi'$ and $\pi^{\ast\ast} = \pi ''$. 
Otherwise, since $\pi'$ is an extreme point of $\text{conv} (\Pi_2 \cup \{ \pi' \}) ~ ( = \text{conv} (\Pi_2 \cup \Pi_1))$ and $\Pi_1$ is not a singleton, there exists $\pi''' \in \Pi_1 \backslash \{ \pi' \}$ such that $\pi' \notin \text{conv} (\Pi_2 \cup \{ \pi''' \}) $. 
Then, we set $\pi^\ast = \pi'''$ and $\pi^{\ast\ast} = \pi'$.)

By Proposition \ref{prop:quasi_hyperplane} and the density of $\mathrm{cone}(\Phi_{\mathbb{F}})+\mathbb{R}$ in $\Phi$, there exist $F, G\in \mathbb{F}$ such that for all $\pi \in  \Pi_2 \cup \{ \pi^{\ast\ast} \}$, 
\begin{equation}
    \langle \varphi_F, \pi^\ast\rangle > \langle \varphi_G, \pi^\ast \rangle ~~~\text{and} ~~~ \langle \varphi_G, \pi \rangle > \langle \varphi_F, \pi \rangle . 
\end{equation}
Therefore, we have $G \not\succsim_1 F$ and $G\succsim_2 F$. Without loss of generality, we assume that for all $f\in F\cup G$ and $\omega \in \Omega$, $f(\omega) \succ x_0$. 
Let $\alpha \in (0,1)$ be such that 
\begin{equation*}
    \langle \varphi_{\alpha G + (1 - \alpha)x_0}, \pi^{\ast\ast} \rangle > \langle \varphi_F, \pi^{\ast\ast} \rangle, 
\end{equation*}
that is, $ F \not \succsim_1  \alpha G + (1 - \alpha)x_0 $. 
By $G \triangleright_D \alpha G + (1 - \alpha)x_0 $ and (iii), $G \not\succsim_2 F$, which is a contradiction to $G\succsim_2 F$.

\subsection*{C.4 Proof of Theorem \ref{thm_jml_comp}}

Let $\succsim_1$ and $\succsim_2$ be JML preferences $(u, \Pi_1)$ and $(u, \Pi_2)$, respectively. 
We first prove that (i) implies (ii). Suppose $\Pi_1 \subset \Pi_2$ and let $F,G\in \mathbb{F}$ such that $F\succ_2 G$. Then, by the definition of the JML preferences, $\min_{\pi \in \Pi_2} ( b^u_F (\pi)  - b^u_G (\pi))  > 0$. 
Since $\Pi_1 \subset \Pi_2$, we have 
\begin{equation*}
    \min_{\pi \in \Pi_1} ( b^u_F (\pi)  - b^u_G (\pi))  \geq \min_{\pi \in \Pi_2}( b^u_F (\pi)  - b^u_G (\pi)). 
\end{equation*}
Therefore, $\min_{\pi \in \Pi_1} ( b^u_F (\pi)  - b^u_G (\pi))   >  0$, which means $F\succ_1 G$. 

We then prove that (ii) implies (i). Suppose to the contrary that (ii) holds but there exists $\pi^\ast \in \Pi_1$ such that $\pi^\ast \notin \Pi_2$. By Proposition \ref{prop:quasi_hyperplane} and  the density of $\mathrm{cone}(\Phi_{\mathbb{F}})+\mathbb{R}$ in $\Phi$, there exist $F, G\in \mathbb{F}$ such that for all $\pi \in \Pi_2$, 
\begin{equation*}
    \langle \varphi_F, \pi\rangle > \langle \varphi_G, \pi \rangle ~~~\text{and} ~~~ \langle \varphi_G, \pi^\ast \rangle > \langle \varphi_F, \pi^\ast \rangle . 
\end{equation*}
Thus, $F\succ_2 G$ and $F\not\succ_1 G$. This is a contradiction to (ii). 

\vspace{5mm}

Next, we show the latter part. Suppose that $\Pi$ is not a singleton. 
We prove that  (i) implies (iii). 
Let $F, G, H\in \mathbb{F}$ such that  $H \triangleright_D F$ and $F \succsim_1 G \succsim_1 H$. 
Then, since $\Pi_1 \subset \Pi_2$ and $F \succsim_1 G \succsim_1 H$, 
    \begin{equation*}
        \max_{\pi \in \Pi_2} (b^u_F (\pi)  - b^u_G (\pi)) 
        \geq \max_{\pi \in \Pi_1} (b^u_F (\pi) - b^u_G (\pi)) \geq 0. 
    \end{equation*}
    and 
    \begin{equation*}
        \max_{\pi \in \Pi_2} ( b^u_G (\pi) -  b^u_H (\pi)) 
        \geq \max_{\pi \in \Pi_1} ( b^u_G (\pi) -  b^u_H (\pi)) \geq 0. 
    \end{equation*}
Therefore, we have $F \succsim_2 G \succsim_2 H$.

    We then prove that (iii) implies (i). Suppose to the contrary that $\Pi_1 \not\subset \Pi_2$. 
    Then, there exist $\pi^\ast, \pi^{\ast\ast} \in \Pi_1$ such that $\pi^\ast \notin \Pi_2$ and $\pi^{\ast\ast} \notin \text{conv} (\Pi_2 \cup \{ \pi^\ast \})$. 
    (Indeed, by $\Pi_1 \not\subset \Pi_2$, there exists $\pi' \in \Pi_1 \backslash \Pi_2$. If there exists $\pi'' $ such that $\pi'' \notin \text{conv} (\Pi_2 \cup \{ \pi' \})$, then define $\pi^\ast$ and $\pi^{\ast\ast}$ as $\pi^\ast = \pi'$ and $\pi^{\ast\ast} = \pi ''$. 
    Otherwise, since $\pi'$ is an extreme point of $\text{conv} (\Pi_2 \cup \{ \pi' \}) ~ ( = \text{conv} (\Pi_2 \cup \Pi_1))$ and $\Pi_1$ is not a singleton, there exists $\pi''' \in \Pi_1 \backslash \{ \pi' \}$ such that $\pi' \notin \text{conv} (\Pi_2 \cup \{ \pi''' \}) $. Then, we set $\pi^\ast = \pi'''$ and $\pi^{\ast\ast} = \pi'$.)

    By proposition \ref{prop:quasi_hyperplane} and the density of $\mathrm{cone}(\Phi_{\mathbb{F}})+\mathbb{R}$ in $\Phi$, there exist $F, G\in \mathbb{F}$ such that for all $\pi \in  \Pi_2 \cup \{ \pi^{\ast\ast} \}$, 
    \begin{equation}
        \langle \varphi_F, \pi^\ast\rangle > \langle \varphi_G, \pi^\ast \rangle ~~~\text{and} ~~~ \langle \varphi_G, \pi \rangle > \langle \varphi_F, \pi \rangle . 
    \end{equation}
    Therefore, we have $F\succsim_1 G$ and $G\succ_2 F$. Without loss of generality, we assume that for all $f\in F\cup G$ and $\omega \in \Omega$, $f(\omega) \succ x_0$. 
    Let $\alpha \in (0,1)$ be such that 
    \begin{equation*}
        \langle \varphi_{\alpha G + (1 - \alpha)x_0}, \pi^{\ast\ast} \rangle > \langle \varphi_F, \pi^{\ast\ast} \rangle, 
    \end{equation*}
    that is, $\alpha G + (1 - \alpha)x_0 \succsim_1 F$. 
    By $G \triangleright_D \alpha G + (1 - \alpha)x_0 $ and (iii), $F\succsim_2 G$, which is a contradiction to $G\succ_2 F$.

\subsection*{C.5 Proof of Theorem \ref{thm:HML}}

Before proving the theorem, we introduce the Vietoris topology, 
which makes our proof more concise. 
Let $(Y, \tau)$ be a topological space. 
We consider the set of  nonempty compact subsets of $Y$, denoted by $\mathcal{K} (Y)$. 
For any finite collection $\{ U, V_1 ,V_2, \cdots, V_n \} \subset \tau$,
let $\text{B} (U ; V_1 ,V_2, \cdots, V_n )$ be a collection of subsets defined as follows: 
For any $W \in \mathcal{K} (Y)$, $W \in \text{B} (U ; V_1 ,V_2, \cdots, V_n )$ if and only if $W\subset U$ and $W\cap V_i \neq \emptyset$ for each $i =1,2, \cdots, n$. 
These sets form a base for a topology on $\mathcal{K} (Y)$. This topology is called the \textit{Vietoris topology}. 

It is known that if $(Y, \tau)$ is metrizable, then its Hausdorff topology coincides with the Vietoris topology in $\mathcal{K} (Y)$ (cf. Theorem 3.91 of \citet{AB2006Math}). 
By using this property, for metrizable topological space $(Y, \tau)$, to prove a function on $\mathcal{K} (Y)$ is continuous in the Hausdorff topology, it is sufficient to check whether this function is continuous in the Vietoris topology or not. 

\vspace{5mm}

Then we prove the only-if part of Theorem \ref{thm:HML}. We omit a proof of the if part since it is straightforward to verify. 
Let $\succsim$ be a binary relation that satisfies \textit{basic axioms}, \textit{completeness for lotteries},  \textit{unambiguous transitivity},  and \textit{reflexivity}. 
Note that we can use all of the results in Appendix B. 
By Claim \ref{claim:K_closed}, the set $K$ is a closed cone including $\{\varphi\in\widetilde{\Phi}_{\mathbb{F}}:\varphi\geq 0\}$.

We show that $K$ can be written as  a union of nonempty convex closed cones including $\{\varphi\in\widetilde{\Phi}_{\mathbb{F}}:\varphi\geq 0\}$.
Take any $\phi\in K\setminus\{\varphi\in\widetilde{\Phi}_{\mathbb{F}}:\varphi\geq 0\}$ and $\phi'\in\{\varphi\in\widetilde{\Phi}_{\mathbb{F}}:\varphi\geq 0\}$.
Then, for any $\alpha\in(0,1)$, $\alpha\phi+(1-\alpha)\phi'\geq\alpha\phi\succsim^\ast 0$.
By the same argument as the latter part of  the proof of Claim \ref{claim:K_closed}, this implies that $\alpha\phi+(1-\alpha)\phi'\succsim^\ast 0$.
Thus,
\begin{equation*}
    K_{\phi}:=\mathrm{co}(\{\alpha\phi:\alpha>0\}\cup\{\varphi\in\widetilde{\Phi}_{\mathbb{F}}:\varphi\geq 0\})
\end{equation*} is a nonempty closed convex cone contained in $K$. 
By construction, we have $K=\cup_{\phi\in K}K_{\phi}$.

Let $\overline{\mathcal{C}}$ be the collection of all non-empty, convex and closed cones that is contained in $K$ and
contains $\{\varphi\in\widetilde{\Phi}_{\mathbb{F}}:\varphi\geq 0\}$. 
Let $\mathcal{C}$ be the subset of $\overline{\mathcal{C}}$ that includes all maximal elements with respect to inclusion.
For each element in $\mathcal{C}$, we can apply the arguments in the proof of Theorem \ref{thm:BML}.
That is, for each $C\in\mathcal{C}$, there exists a nonempty closed convex set $\Pi_{C}$ of probability distributions over $\Delta(\Omega)$ such that $\varphi\in C$ if and only if $\langle\varphi,\pi\rangle\geq0$ for all $\pi\in\Pi_{C}$.
Denote $\mathbf{\Pi}^\circ=\{\Pi_{C}\}_{C\in\mathcal{C}}$.

\begin{claim}
\label{claim:loosely-closed}
    For all $\bar{\Pi}\in \text{cl} (\mathbf{\Pi}^\circ)$, there exists  $\Pi\in\mathbf{\Pi}^\circ$ such that $\Pi\subset \bar{\Pi}$.
\end{claim}

\begin{proof}
For each $\Pi'\in \mathrm{cl}(\mathbf{\Pi}^\circ)$, we define the cone $C_{\Pi'}$ as 
\begin{equation*}
    C_{\Pi'}:=\{\varphi\in\tilde{\Phi}_{\mathbb{F}}:\langle\varphi,\pi\rangle\geq 0\text{ for all }\pi\in\Pi'\}.
\end{equation*}
By construction, $C_{\Pi'}$ corresponds one to one with $\Pi'$.

Suppose to the contrary that there exists $\bar{\Pi}\in \text{cl} (\mathbf{\Pi}^\circ)$ such that for all $\Pi\in\mathbf{\Pi}^\circ$, $\Pi\not\subset\bar{\Pi}$.
If $C_{\bar{\Pi}} \subset K$, then by the constructions of $\mathcal{C}$ and $\mathbf{\Pi}^\circ$, there exists $\Pi \in \mathbf{\Pi}^\circ$ such that $C_{\bar{\Pi}} \subset C_{\Pi}$, which is equivalent to $\Pi \subset \bar{\Pi}$. 
This is a contradiction to the assumption that  for all $\Pi\in\mathbf{\Pi}^\circ$, $\Pi\not\subset\bar{\Pi}$.
Thus, there exists $\varphi \in C_{\bar{\Pi}}$ such that $\varphi \notin K$. 

Since $\bar{\Pi}\in \text{cl} (\mathbf{\Pi}^\circ)$, there exists a sequence  $\{\Pi_{k}\}_{k\in\mathbb{N}}\subset\mathbf{\Pi}^\circ$ that converges to $\bar{\Pi}$ in the Hausdorff topology.
We define a distance between two cones in $\{C_{\Pi_k}\}_{k\in \mathbb{N}}$ by the Hausdorff metric between two corresponding two probability sets in $\{\Pi_k\}_{k\in \mathbb{N}}$.
This is well defined since each $C_{\Pi_k}$ corresponds one to one with $\Pi_k$.\footnote{The corresponding discussion was presented in footnote 16 of \citet{lehrer2011justifiable}.}
Then, the sequence $\{C_{\Pi_k}\}_{k\in \mathbb{N}}$ converges to $C_{\bar{\Pi}}$. 
Therefore, there exists a sequence $\{\varphi_k\}_{k\in \mathbb{N}}$ such that $\varphi_k \in C_{\Pi_k}\subset K$ for all $k\in\mathbb{N}$ and $\varphi_k \rightarrow\varphi$.
This contradicts the fact that $K$ is closed.
\end{proof}

We next show that for all $\varphi \in \widetilde{\Phi}_\mathbb{F}$, 
\begin{equation}
\label{eq:dualself_utility}
    \varphi\in K\iff\max_{\Pi\in\mathbf{\Pi}^\circ}\min_{\pi\in\Pi}\langle\varphi,\pi\rangle\geq 0.
\end{equation}
To prove this, it is sufficient to show that there is no $\varphi \in \widetilde{\Phi}_{\mathbb{F}}$ such that ${\min_{\pi\in\Pi}\langle\varphi,\pi\rangle} < 0$ for all $\Pi \in \mathbf{\Pi}^\circ$ but $\sup_{\Pi\in\mathbf{\Pi}^\circ}\min_{\pi\in\Pi}\langle\varphi,\pi\rangle = 0$. 
Suppose to the contrary that for some $\varphi^+ \in \widetilde{\Phi}_{\mathbb{F}}$, ${\min_{\pi\in\Pi}\langle\varphi^+,\pi\rangle} < 0$ for all $\Pi \in \mathbf{\Pi}^\circ$ but $\sup_{\Pi\in\mathbf{\Pi}^\circ}\min_{\pi\in\Pi}\langle\varphi^+,\pi\rangle = 0$. 

For $\varphi \in \widetilde{\Phi}_{\mathbb{F}}$,  define $G_\varphi : \mathcal{K}(\Delta (\Delta (\Omega))) \rightarrow \mathbb{R}$ as for all $\Pi \in \mathcal{K}(\Delta (\Delta (\Omega))) $, $G_\varphi (\Pi) = \min_{\pi \in \Pi} \langle\varphi,\pi\rangle$. Note that the minimum can be  achieved since $\Pi \in \mathcal{K}(\Delta (\Delta (\Omega))) $ is closed.
We prove a preliminary result. 

\begin{claim}
\label{claim:cont-G_varphi}
    The function $G_\varphi$ is continuous in the Hausdorff topology. 
\end{claim}

\begin{proof}
    Take $\Pi \in \mathcal{K}(\Delta (\Delta (\Omega)))$ and $\varepsilon > 0$ arbitrarily. 
    Let $\Pi_1, \Pi_2$ be the open sets of $\Delta (\Delta (\Omega))$ such that 
    \begin{align*}
        \Pi_1 &= \{ \pi \in \Delta (\Delta (\Omega)) : | \langle\varphi,\pi\rangle -  G_\varphi (\Pi) | < \varepsilon \}, 
        \\
        \Pi_2 &= \qty{ \pi \in \Delta (\Delta (\Omega)) : \langle\varphi,\pi\rangle >  G_\varphi (\Pi) - {1 \over 2} \varepsilon }.  
    \end{align*}
    Note that  $\Pi \in \text{B} (\Pi_1 \cup\Pi_2; \Pi_1, \Pi_2)$. 
    Then, for all $\Pi' \in \text{B} (\Pi_1 \cup\Pi_2; \Pi_1, \Pi_2)$, we have $| G_\varphi (\Pi') - G_\varphi (\Pi)|  = | \min_{\pi \in \Pi'} \langle\varphi,\pi\rangle - G_\varphi (\Pi) | < \varepsilon$. 
    Since $ \text{B} (\Pi_1 \cup\Pi_2; \Pi_1, \Pi_2)$ is a open set including  $\Pi$ in the Vietoris topology and the Vietoris topology coincides with the Hausdorff topology in $\mathcal{K} (\Delta (\Delta (\Omega)))$, $G_\varphi$ is continuous in the Hausdorff topology. 
\end{proof}

Together with ${\min_{\pi\in\Pi}\langle\varphi^+,\pi\rangle} < 0$ for all $\Pi \in \mathbf{\Pi}^\circ$ and $\sup_{\Pi\in\mathbf{\Pi}^\circ}\min_{\pi\in\Pi}\langle\varphi^+,\pi\rangle = 0$,  Claim \ref{claim:cont-G_varphi} implies that there exists  $\Pi^\ast \in \text{cl} (\mathbf{\Pi}^\circ)$ such that $\min_{\pi\in\Pi^\ast}\langle\varphi^+,\pi\rangle = 0$. 
By Claim \ref{claim:loosely-closed}, there exists  $\Pi'\in\mathbf{\Pi}^\circ$ with $\Pi' \subset \Pi^\ast$.
Therefore, we have $\min_{\pi\in\Pi'}\langle\varphi^+,\pi\rangle \geq \min_{\pi\in\Pi^\ast}\langle\varphi^+,\pi\rangle =  0$.
This is a contradiction to the assumption that ${\min_{\pi\in\Pi}\langle\varphi^+,\pi\rangle} < 0$ for all $\Pi \in \mathbf{\Pi}^\circ$. 
Thus, we have \eqref{eq:dualself_utility}.

Let $\mathbf{\Pi} = \text{cl} (\mathbf{\Pi}^\circ)$. 
Since $\mathcal{K} (\Delta (\Delta (\Omega)))$ is a compact space (cf. Theorem 15.15 and 3.88 of \citet{AB2006Math}), $\mathbf{\Pi}$ is also compact. 
Since $G_\varphi$ is continuous for any $\varphi \in \widetilde{\Phi}_\mathbb{F}$ (Claim \ref{claim:cont-G_varphi}), we have 
\begin{equation*}
    \max_{\Pi \in \mathbf{\Pi}} \min_{\pi \in \Pi} \langle\varphi,\pi\rangle = \max_{\Pi \in \mathbf{\Pi}} G_\varphi (\Pi) = \max_{\Pi \in \mathbf{\Pi}^\circ} G_\varphi (\Pi) = \max_{\Pi \in \mathbf{\Pi}^\circ} \min_{\pi \in \Pi} \langle\varphi,\pi\rangle. 
\end{equation*}
Therefore, by \eqref{eq:dualself_utility}, we obtain 
\begin{equation*}
    \varphi\in K\iff\max_{\Pi\in\mathbf{\Pi}}\min_{\pi\in\Pi}\langle\varphi,\pi\rangle\geq 0.
\end{equation*}




\subsection*{C.6 Proof of Theorem \ref{thm_alpha_rat}}

This section proves that (i) implies (ii). The converse is straightforward to verify.  
Let $\succsim$ be an HML preference $(u, \mathbf{\Pi})$ and $\succsim^\land$ be a binary relation that satisfies \textit{completeness}, \textit{transitivity}, and \textit{mixture continuity}. Furthermore, suppose that the pair $(\succsim, \succsim^\land)$ satisfies \textit{lottery consistency} and \textit{robustly strict consistency}. 

    Take $x,y\in X$ arbitrary. If $x\succ y$, then $x\succ {1\over 2}x + {1\over 2}y \succ y$. 
    By the definition of $\stsucc$, we have $x\stsucc {1\over 2}x + {1\over 2}y \stsucc y$. By \textit{robustly strict consistency}, we obtain $x\succ^\land y$. If $x\succsim y$, then by \textit{lottery consistency}, $x\succsim^\land y$. 
    Therefore, 
    \begin{equation}
    \label{eq:u_eq2}
        x\succsim^\land y \iff x\succsim y \iff u(x) \geq u(y). 
    \end{equation}

\begin{claim}
    For all $F\in \mathbb{F}$, there exists $x_F \in X$ such that $F\sim^\land x_F$. 
\end{claim}

\begin{proof}
    Let $x^\ast, x_\ast \in X$ be such that $x^\ast \succ^\land x_\ast$ and $x^\ast \succsim^\land f(\omega) \succsim^\land x_\ast$ for all $f\in F$ and $\omega \in \Omega$.\footnote{Since $\succsim$ satisfies \textit{nontriviality} and both of $\Omega$ and $F$ are finite, such a pair $(x^\ast, x_\ast)$ exists. }
    We verify that $x^\ast \succsim^\land F \succsim^\land x_\ast$. 
    Let $\alpha \in (0,1)$. 
    By \eqref{eq:u_eq2},  $x^\ast \succ \alpha f (\omega) + (1-\alpha) x_\ast $ for all $f\in F$, and $\omega \in \Omega$. 
    Furthermore, since $u$ is affine, there exists $y \in \mathrm{co}\{ x^\ast, x_\ast\}$ such that $x^\ast \succ y \succ \alpha f (\omega) + (1-\alpha) x_\ast $ for all $f\in F$ and $\omega \in \Omega$. Since $\succsim$ is an HML preference, we have  $x^\ast \stsucc y \stsucc \alpha F + (1-\alpha) x_\ast $. By \textit{robustly strict consistency}, $x^\ast \succ^\land \alpha F + (1-\alpha) x_\ast $. By \textit{mixture continuity}, we have $x^\ast \succsim^\land F$. 

    To prove $F \succsim^\land x_\ast$, note that for all $\alpha' \in (0,1)$, there exists $y'\in  \mathrm{co}\{ x^\ast, x_\ast\}$ such that $\alpha' f (\omega) + (1-\alpha') x^\ast \succ y' \succ x_\ast$ for all $f\in F$ and $\omega \in \Omega$. 
    Since $\succsim$ is an HML preference, we have  $\alpha' F + (1-\alpha') x^\ast \stsucc y' \stsucc x_\ast$. 
    By \textit{robustly strict consistency}, $\alpha' F + (1-\alpha') x^\ast \succ^\land x_\ast$.
    By \textit{mixture continuity}, we have $F \succsim^\land x_\ast$. 
    
   Then by \textit{mixture continuity}, the nonempty sets 
   \begin{equation*}
       A = \{ \beta \in [0,1] : \beta x^\ast + (1 - \beta ) x_\ast \succsim^\land F \} ~~~ \text{and} ~~~ B = \{ \beta \in [0,1] : F \succsim^\land  \beta x^\ast + (1 - \beta ) x_\ast  \}
   \end{equation*}
   are closed and $A\cup B = [0,1]$. Therefore, there exists $\tilde{\beta} \in A \cap B$. Let $x_F = \tilde{\beta} x^\ast + (1 - \tilde{\beta} ) x_\ast$. Then, $x_F\sim^\land F$, as required.
\end{proof}

    Take $F\in\mathbb{F}$ arbitrarily.
    Let 
    \begin{align*}
        \underline{\gamma} &= \min \qty{ \max_{\Pi \in \mathbf{\Pi}} \min_{\pi \in \Pi} b_F^u (\pi) , \min_{\Pi \in \mathbf{\Pi}} \max_{\pi \in \Pi} b_F^u (\pi)  }, \\
        \overline{\gamma} &= \max \qty{ \max_{\Pi \in \mathbf{\Pi}} \min_{\pi \in \Pi} b_F^u (\pi) , \min_{\Pi \in \mathbf{\Pi}} \max_{\pi \in \Pi} b_F^u (\pi)  }. 
    \end{align*}
    We prove that $\underline{\gamma} \leq u(x_F) \leq \overline{\gamma}$. 
    If $u(x_F) < \underline{\gamma}$, then there exists $y\in X$ such that $u(x_F) < u(y) < \underline{\gamma}$. 
    Then, by the definition of $\stsucc$, we have $F\stsucc y \stsucc x_F$. By \textit{robustly strict consistency}, $F \succ^\land x_F$, which is a contradiction to the definition of $x_F$. Thus, $\underline{\gamma} \leq u(x_F) $ holds. 
    In the same way, we can prove $ u(x_F) \leq \overline{\gamma}$. 

    Therefore, for each $F \in \mathbb{F}$, there exists $\beta_F \in [0,1]$ such that 
    \begin{equation*}
        u(x_F) = \beta_F \underline{\gamma} + (1 - \beta_F)\overline{\gamma}. 
    \end{equation*}
    Then, we define the function $\alpha : \mathbb{F} \rightarrow [0,1]$ such that for all $F\in \mathbb{F}$, 
    \begin{equation*}
        \alpha (F)
        = 
            \left\{
            \begin{array}{ll}
            \beta_F &  \text{if} ~~ \max_{\Pi \in \mathbf{\Pi}} \min_{\pi \in \Pi} b_F^u (\pi) \leq  \min_{\Pi \in \mathbf{\Pi}} \max_{\pi \in \Pi} b_F^u (\pi),  \\
            (1-\beta_F) & \text{if} ~~ \max_{\Pi \in \mathbf{\Pi}} \min_{\pi \in \Pi} b_F^u (\pi) > \min_{\Pi \in \mathbf{\Pi}} \max_{\pi \in \Pi} b_F^u (\pi) . 
            \end{array}
            \right.
    \end{equation*}
    Then, the function $U:\mathbb{F} \to \mathbb{R}$ defined as  for all $F\in \mathbb{F}$, 
     \begin{equation*}
            U(F) = \alpha (F) \max_{\Pi \in \mathbf{\Pi}} \min_{\pi \in \Pi} b_F^u (\pi)  + (1 - \alpha (F)) \min_{\Pi \in \mathbf{\Pi}} \max_{\pi \in \Pi} b_F^u (\pi) 
        \end{equation*} 
    represents $\succsim^\land$. 

\bibliographystyle{econ}
\bibliography{reference}

\end{document}